\documentclass[fleqn,usenatbib]{mnras}

\usepackage{newtxtext,newtxmath}
\usepackage[T1]{fontenc}

\DeclareRobustCommand{\VAN}[3]{#2}
\let\VANthebibliography\thebibliography
\def\thebibliography{\DeclareRobustCommand{\VAN}[3]{##3}\VANthebibliography}

\usepackage{graphicx}
\usepackage{amsmath}
\usepackage{xcolor}
\usepackage{multirow}
\usepackage{subcaption}

\title[$S_8$ Tension and Interacting Dark Sectors]{\boldmath Reconciling $S_8$: Insights from Interacting Dark Sectors}

\author[R. Shah, P. Mukherjee, S. Pal]{
Rahul Shah,$^{1}$\thanks{E-mail: rahul.shah.13.97@gmail.com}
Purba Mukherjee,$^{1,2}$\thanks{E-mail: purba16@gmail.com}
and Supratik Pal$^{1,3}$\thanks{E-mail: supratik@isical.ac.in}
\\
$^{1}$Physics and Applied Mathematics Unit, Indian Statistical Institute, 203 B.T. Road, Kolkata 700 108, India\\
$^{2}$Centre for Theoretical Physics, Jamia Millia Islamia, New Delhi 110025, India\\
$^{3}$Technology Innovation Hub on Data Science, Big Data Analytics and Data Curation, Indian Statistical Institute, 203 B.T. Road, Kolkata 700 108, India
}

\begin{document}
\label{firstpage}
\pagerange{\pageref{firstpage}--\pageref{lastpage}}
\maketitle

\begin{abstract}
We do a careful investigation of the prospects of dark energy (DE) interacting with cold dark matter in alleviating the $S_8$ clustering tension. To this end, we consider various well-known parametrizations of the DE equation of state (EoS) and consider perturbations in both the dark sectors, along with an interaction term. Moreover, we perform a separate study for the phantom and non-phantom regimes. Using cosmic microwave background (CMB), baryon acoustic oscillations, and Type Ia supernovae data sets, constraints on the model parameters for each case have been obtained and a generic reduction in the $H_0-\sigma_{8,0}$ correlation has been observed, both for constant and dynamical DE EoS. This reduction, coupled with a significant negative correlation between the interaction term and $\sigma_{8,0}$, contributes to easing the clustering tension by lowering $\sigma_{8,0}$ to somewhere in between the early CMB and late-time clustering measurements for the phantom regime, for almost all the models under consideration. Additionally, this is achieved without exacerbating the Hubble tension. In this regard, the interacting Chevallier–Polarski–Linder and Jassal-Bagla-Padmanabhan models perform the best in relaxing the $S_8$ tension to $<1\sigma$. However, for the non-phantom regime the $\sigma_{8,0}$ tension tends to have worsened, which reassures the merits of phantom DE from latest data. We further investigate the role of redshift space distortion data sets and find an overall reduction in tension, with a $\sigma_{8,0}$ value relatively closer to the CMB value. We finally check whether further extensions of this scenario, such as the inclusion of the sound speed of DE and warm dark matter interacting with DE, can have some effects.
\end{abstract}

\begin{keywords}
methods: data analysis -- methods: statistical -- cosmological parameters -- dark energy -- cosmology: observations -- cosmology: theory 
\vspace{-7pt}
\end{keywords}

\section{Introduction}\label{sec:introduction}
The $\Lambda$ cold dark matter ($\Lambda$CDM) cosmological model, which assumes dark energy (DE) as a cosmological constant ($\Lambda$) and non-relativistic ``cold'' dark matter (CDM) as the primary constituents of the Universe, has provided valuable insights into various astrophysical and cosmological phenomena. However, recent inconsistencies arising from precise measurements of the model parameters, from different scales of observations, have raised questions regarding its effectiveness. Chief among these concerns is the 5$\sigma$ Hubble tension \citep{Hazra:2013dsx,Novosyadlyj:2013nya,deg2}, which has garnered significant attention in the scientific community, leading to numerous proposed explanations and alternative theoretical models \citep{DiValentino:2021izs,H0Olympics,snowmasstensions,Vagnozzi:2023nrq,DAgostino:2023cgx}. However, a different, yet closely related tension exists in the measurements of the matter clustering strength within the Universe, as quantified by the $\sigma_{8,0}$ or $S_8=\sigma_{8,0}\sqrt{\Omega_{m0}/0.3}$ parameter \citep{DiValentino:2020vvd}. A discrepancy of $\sim3\sigma$ exists between the values derived from cosmic microwave background (CMB) data from \textit{Planck} 2018 \citep{Pl18VI} ($S_8=0.832\pm0.013/\sigma_{8,0}=0.8111\pm0.0060$) and those inferred from observations at lower redshifts, viz. Canada-France-Hawaii Telescope Lensing Survey (CFHTLenS) \citep{Joudaki:2016mvz} ($S_8=0.732^{+0.029}_{-0.031}$); Kilo Degree Survey 450 (KiDS-450) \citep{Joudaki:2016kym} ($S_8=0.745\pm0.039$), which when combined with 2 degree Field Lensing Survey (2dFLenS) \citep{Joudaki:2017zdt} gives $S_8=0.742\pm0.035$; KiDS+VIKING-450 (KV450) \citep{Hildebrandt:2018yau} ($S_8=0.737^{+0.040}_{-0.036}$), on adding Baryon Oscillation Spectroscopic Survey (BOSS) \citep{Troster:2019ean} gives $S_8=0.728\pm0.026$. Dark Energy Survey Year 1 (DES-Y1) \citep{DES:2017qwj} obtains $S_8=0.782\pm0.027$, and in combination with KV450 \citep{Joudaki:2019pmv} one gets $S_8=0.755^{+0.019}_{-0.021}$. KiDS-1000 \citep{KiDS:2020suj} alone gives $S_8=0.759^{+0.024}_{-0.021}$, and KiDS-1000+BOSS+2dFLenS \citep{Heymans:2020gsg} finds $S_8=0.766^{+0.020}_{-0.014}$. DES-Y3 \citep{DES:2021bvc} finds $S_8=0.759^{+0.025}_{-0.023}$. However, at the same time, a few late-time probes give constraints closer to the early-time measurements, such as those from KiDS-450+Galaxies And Mass Assembly (GAMA) \citep{vanUitert:2017ieu} ($S_8=0.800^{+0.029}_{-0.027}$) and Hyper Suprime-Cam Subaru Strategic Program (HSC SSP) \citep{Hamana:2019etx} ($S_8=0.804^{+0.032}_{-0.029}$).

Although the $\Lambda$CDM model more or less fits either set of data, observations at lower redshifts, in general, unarguably suggest a reduced level of structure formation resulting from the discrepancy in the inferred values of $S_8$. Since the status of the clustering tension is less apparent than that of the Hubble tension, investigations into this aspect are significantly overshadowed by studies on the latter. However, that there exists a strong positive correlation between $H_0$ and $\sigma_{8,0}$ has been pointed out earlier by \citet{Bhattacharyya:2018fwb}. Some studies suggest a possible connection between these tensions through the $\Omega_{m0}$ parameter \citep{Akarsu:2024qiq,Colgain:2022nlb,Adil:2023jtu,Colgain:2024xqj}. Treatments to alleviate one of these tensions often exacerbate the other, highlighting the need for a joint approach to address both consistently. However, addressing the $S_8$ tension is more complex to consider, both theoretically and observationally, as the $\sigma_{8,0}$ parameter is particularly sensitive to cosmological dynamics at the perturbative level, compared to the background evolution dependent $H_0$. Herein lies the primary challenge in today's precision cosmology - to simultaneously address both tensions, operating at different levels of observation, or at the very least, to address one without compromising the other.

There have been some level of efforts to alleviate the clustering tension \citep{snowmasstensions,DiValentino:2020vvd}. In this study, we aim to reassess this tension within the framework of interacting dark matter and dark energy (iDMDE) models, which entails an exchange of energy between the dark sectors. Initially proposed to solve the coincidence problem in cosmology \citep{Amendola:1999er}, these models have lately attracted a lot of attention in addressing the Hubble tension \citep{Giare:2024ytc,Benisty:2024lmj,Hoerning:2023hks,Lucca:2020zjb,Vagnozzi:2023nrq,DiValentino:2017iww,DiValentino:2019ffd,Pan:2023mie,Wang:2024vmw,Yang:2018euj,Bhattacharyya:2018fwb,Bernui:2023byc,Yao:2022kub,Gariazzo:2021qtg,Guo:2021rrz,Nunes:2021zzi,Zhao:2022ycr,Gao:2021xnk,Pan:2020bur,Amirhashchi:2020qep,Pan:2019gop,Pan:2019jqh,Yang:2018uae,Gao:2022ahg,Kumar:2019wfs,Kumar:2021eev,Yang:2021hxg}. Of late, they have been explored in the context of the clustering tension as well. For example, \citet{DiValentino:2019ffd,DiValentino:2019jae} consider an iDMDE set-up for $\Lambda$CDM, imposing hard bounds on the equation-of-state (EoS) parameter to avoid instabilities in the perturbations (also see \citet{Lucca:2021dxo}). In \citet{Bhattacharyya:2018fwb}, an iDMDE treatment with perturbations is done by absorbing the interaction in terms of effective parameters, considering a Chevallier–Polarski–Linder (CPL) \citep{cpl1} parametrization of the effective EoS of DE. As a result, the correlation between the interaction term and other parameters is not evident. \citet{Gao:2022ahg, Kumar:2019wfs, Kumar:2021eev} consider interactions, but within the $\Lambda$CDM framework only, with no perturbations in the DE sector. \citet{An:2017crg} consider various forms of the interaction term and obtain constraints on the clustering parameter by keeping the EoS of DE free, but constant, \textit{i.e.} an interacting $w$ cold dark matter ($w$CDM) set-up. Similar efforts along these directions were carried out in \citet{Mukherjee:2017oom, Sinha:2019axe, Sinha:2021tnr, Sinha:2022dze} for different forms of the interaction term or alternative DE models. Usually, an interacting set-up is characterized primarily by the coupling term, which governs the magnitude and direction of energy-momentum transfer. However, consensus regarding the precise form of this interaction remains elusive due to the inherently ``dark'' nature of the involved components, which hinders a well-founded first-principles field-theoretic formulation.

In this study, we aim to revisit and expand upon the existing literature by doing a systematic case-by-case analysis focusing on the clustering tension, albeit with the latest compilation of data sets, thus improving upon some of the existing studies, which have since become outdated. Here, we utilize a combination of the latest CMB data from \textit{Planck} 2018 \citep{Pl18VI}, baryon acoustic oscillations (BAO) data, and Type Ia supernovae (SNIa) observations from the Pantheon+ \citep{panprelease} compilation. We begin by considering an interaction term proportional to the DE density, a framework that has exhibited some promise in addressing cosmological tensions \citep{Hoerning:2023hks,Lucca:2020zjb,DiValentino:2017iww,DiValentino:2019ffd,Liu:2022hpz,Wang:2016lxa,Yang:2019uog,Mukhopadhyay:2020bml,Bamba:2012cp}. Unlike some of the previous approaches using effective parameters, we adopt a generic prescription and explore this model under different parametrizations of the DE EoS. This exercise can in turn help us explore whether these extensions to $\Lambda$ can alleviate tensions in the presence of interactions. Nevertheless, we keep the provision open so that both DM and DE can take part in clustering, by considering generic first order cosmological perturbations in both the DM and DE sectors, thereby carefully accounting for the effect of the interactions and the nature of DE on the perturbative level.

It is crucial to highlight that, when perturbation in DE is considered, the velocity perturbation equation for the DE sector blows up if the EoS reaches exactly $w=-1$ at any point during cosmic evolution. Thus, for the dynamical DE (DDE) models, it is imperative to prevent phantom-crossing to avoid instabilities in the perturbation equations. Hence we deal with the phantom and non-phantom regimes separately as they are entirely motivated by fundamentally distinct physics. This is the only restriction we impose in our analysis, thereby refraining from imposing any additional hard bounds on the interaction parameter or any other parameter, as has been commonly done in the past to avoid theoretical instabilities due to interactions. We keep the interaction parameter entirely free to explicitly examine whether such instabilities arise in light of current data sets, hence enabling generic conclusions. We find that the presence of interactions for a phantom EoS of DE helps relax the $\sigma_{8,0}$ tension, without exacerbating the $H_0$ tension. However, a non-phantom case worsens the situation and hence is not favoured by latest data sets. We also see a generic reduction, and in some cases, a complete elimination of the $H_0-\sigma_{8,0}$ correlation for the interacting constant/DDE (i-C/DDE) parametrizations. This allows for reduction of the $\sigma_{8,0}$ tension without significant changes in $H_0$.

Since the $\sigma_{8,0}$ parameter is under focus, we further investigate the effect of incorporating the $f\sigma_8$ measurements from redshift space distortion (RSD) \citep{Kaiser:1987qv} observations in addition to the aforementioned data sets. We find noticeably tighter constraints on certain parameters, including $\sigma_{8,0}$, which take relatively higher values compared to non-inclusion of RSD data. Although there still is an overall reduction in the clustering tension, the extent of it is reduced. 

Moreover, the role of DE clustering, however small, can be revealed in the context of tensions, providing valuable insights into the behaviour of DE at small scales \citep{Hannestad:2005ak, Xia:2007km, de_Putter_2010, Linton:2017ged,Dinda:2023mad}. Therefore, attempts were made to constrain the DE sound speed $c_s^2$ by keeping it as a free parameter in our analyses. We found that $c_s^2$ remains almost unconstrained with current data sets, and hence has little effect on other parameters of interest. As a demonstrative case, we show how the overall constraints are affected if it is kept as a free parameter for interacting $w$CDM. Moreover, as a minimal extension to the above scenarios, we investigate the role of a ``non-cold'' DM sector, parametrized via some non-zero but constant DM EoS, described by an extra free parameter $\mathrm{w_{DM}}$. 

We place particular attention to the $\sigma_{8,0}/S_8$ tension with our goal being to examine not only the effect of interactions within the dark sector but also the impact of DE perturbations and different parametrizations of C/DDE explicitly. In section \ref{sec:iDMDE}, we provide a brief outline regarding the theoretical framework of the interacting dark sector. Section \ref{sec:models_datasets} introduces various parametrizations of the EoS of DE in the interacting scenario considered in this work, along with an outline of the data sets employed for this study. Our results and analysis are presented in section \ref{sec:results}. In Section \ref{sec:RSD}, we study the effects of adding RSD data to the analysis. Section \ref{sec:wdm} explores the prospects of warm DM (WDM) in some of the interacting scenarios to gain insights into the role of the nature of DM in this context. Finally, we make some concluding remarks in section \ref{sec:conclusion}.

\section{Interacting two-fluid set-up with CDM and DE}\label{sec:iDMDE}
Understanding the interaction within the dark sector poses a significant challenge due to the lack of a fundamental understanding of each component, unlike the situation in the standard model of particle physics. However, a practical approach followed is to consider both DE and DM as ideal fluids, having a transfer of energy or momentum between both sectors. This perspective circumvents the need for a Lagrangian description of the system to gain cosmological insights. For a non-zero interaction between the two dark sectors, the energy-momentum tensors of DE and DM are not independently conserved; rather,
\begin{equation}
    \nabla_\mu T^{\mu\nu}_{(k)} = \mathcal{Q}^\nu_{(k)}
\end{equation}
is satisfied, where the index $k$ is ``de'' for DE and ``dm'' for DM, with $\mathcal{Q}^\nu_{(k)}$ quantifying the energy-momentum flux between them. Further, the Bianchi identity dictates the overall conservation of the energy-momentum tensor, and hence $\mathcal{Q}^\nu_{\mathrm{de}}=-\mathcal{Q}^\nu_{\mathrm{dm}}$. In this work, we consider only energy transfers to be possible $\left(\mathcal{Q}^0_{\mathrm{de}}=-\mathcal{Q}^0_{\mathrm{dm}} = \mathcal{Q}\right)$, with no momentum transfer, implying $\mathcal{Q}^i_{(k)}=0$.

Assuming DM to be cold, \textit{i.e.} non-relativistic (CDM), we have the continuity equations,
\begin{equation}
    \dot{\rho}_{\mathrm{dm}} + 3\mathcal{H}\rho_{\mathrm{dm}} = a^2 \mathcal{Q}^0_{\mathrm{dm}} = a\mathcal{Q}\:,
\end{equation}
\begin{equation}
    \dot{\rho}_{\mathrm{de}} + 3\mathcal{H}(1 + w)\rho_{\mathrm{de}} = a^2 \mathcal{Q}^0_{\mathrm{de}} = -a\mathcal{Q}\:,
\end{equation}
where an overhead dot represents derivative with respect to conformal time, $\mathcal{H}=aH$ is the conformal Hubble parameter, $w\equiv w(a)$ is the EoS of DE, and $\mathcal{Q}$ is the rate of transfer of energy density, with $\mathcal{Q}>0$ implying an energy transfer from DE to DM [as a solution to the coincidence problem \citep{Olivares:2005tb}], and vice versa.

We assume the form of the interaction to be
\begin{equation}
    \mathcal{Q} = H Q \rho_{\mathrm{de}}\:,
\end{equation}
where $Q$ is the coupling constant. The analytical solutions to the continuity equations above are then
\begin{equation}
    \rho_{\mathrm{dm}}=\rho_{\mathrm{dm}, 0} \, a^{-3} \, + \rho_{\mathrm{de}, 0} \, a^{-3} \, Q \int_1^a \frac{a'^2 \, e^{\xi(a')}}{a'^Q} \, \mathrm{d} a' \:,
\end{equation}
\begin{equation}
    \rho_{\mathrm{de}}=\rho_{\mathrm{de}, 0} \, \frac{e^{\xi(a)}}{a^Q}\:,
\end{equation}
where $\xi(a)=\int_{a}^{1} \frac{3[1+w(a')]}{a'}\mathrm{d} a'$.\\

The linear order perturbation equations for DM and DE in the synchronous gauge are given as \citep{Hoerning:2023hks,Valiviita:2008iv,Gavela:2010tm,DaCosta:2014vsr,deCesare:2021wmk,Ma:1995ey}
\begin{equation}
    \dot{\delta}_{\mathrm{dm}}=-\theta_{\mathrm{dm}}-\frac{\dot{h}}{2}+\mathcal{H} Q \frac{\rho_{\mathrm{de}}}{\rho_{\mathrm{dm}}}\left(\delta_{\mathrm{de}}-\delta_{\mathrm{dm}}\right)+Q \frac{\rho_{\mathrm{de}}}{\rho_{\mathrm{dm}}}\left(\frac{k v_T}{3}+\frac{\dot{h}}{6}\right)\:,
\end{equation}
\begin{equation}
    \dot{\theta}_{\mathrm{dm}}=-\mathcal{H} \theta_{\mathrm{dm}}\left(1+Q \frac{\rho_{\mathrm{de}}}{\rho_{\mathrm{dm}}}\right)\:,
\end{equation}
\begin{align}
    \dot{\delta}_{\mathrm{de}}=&-(1+w)\left(\theta_{\mathrm{de}}+\frac{\dot{h}}{2}\right)- \frac{3\mathcal{H}\theta_{\mathrm{de}}}{k^2}\dot{w} -Q\left(\frac{k v_T}{3}+\frac{\dot{h}}{6}\right) \notag\\
    &-3 \mathcal{H}(c_s^2-w) \left[\delta_{\mathrm{de}}+\{3(1+w)+Q\} \frac{\mathcal{H} \theta_{\mathrm{de}}}{k^2}\right]\:,
\end{align}
\begin{equation}
    \dot{\theta}_{\mathrm{de}}= -\mathcal{H} \left(1-3c_s^2\right) \theta_{\mathrm{de}} + \frac{Q\mathcal{H}\theta_{\mathrm{de}}}{1+w}\left(1+c_s^2\right) +\frac{k^2c_s^2}{1+w} \delta_{\mathrm{de}}\:,
\end{equation}
where $\delta_{\mathrm{dm}}$ and $\delta_{\mathrm{de}}$ denote the density contrasts, $\theta_{\mathrm{dm}}$ and $\theta_{\mathrm{de}}$ denote the velocity perturbations, and $\rho_{\mathrm{dm}}$ and $\rho_{\mathrm{de}}$ denote the densities, of DM and DE, respectively. The DE adiabatic sound speed is set to the EoS of DE ($w$), $c_s^2$ is the sound speed of DE, and $v_T$ is the centre-of-mass velocity for the total fluid defined as $(1+w_T)v_T=\sum_a (1+w_a)\Omega_a v_a$. The extra terms in the density contrast equations are due to the perturbation of the Hubble rate ($\delta H$) \citep{Hoerning:2023hks}. 

\section{DE models and data sets used}\label{sec:models_datasets}
In this study, we characterize DM as non-relativistic (CDM). We explore interacting scenarios by employing several well-established parametrizations of the EoS of DE. Perturbations are incorporated into both dark sectors to enable insights into the combined influences of interaction and perturbations. The diverse parametrizations of the DE EoS under examination are henceforth referred to as different models:
\begin{enumerate}
    \item i-$\Lambda$CDM: We include an interacting $\Lambda$CDM as the benchmark model in our study, where the DE sector is characterized by a constant EoS $w=-1$, and, obviously, it does not take part in perturbations. We have included this in our analysis only to compare the performances of the alternative models against this baseline.
    \item i-$w$CDM: Interacting $w$CDM is a minimal extension of $\Lambda$CDM where the DE EoS $w$ is kept as a constant (not redshift-evolving) but otherwise free parameter, \textit{i.e.} $w(a)=w_0$. 
    \item i-CPL: The CPL parametrization of DE \citep{cpl1,eos1}, which is the most widely used two-parameter extension to $\Lambda$CDM, has a redshift-dependent DE EoS given by
    \begin{equation}
        w(a)=w_0+w_a \, (1-a)\:\:.
    \end{equation}
    The present value of the EoS is given by $w_0$, while it remains bounded by $w_0+w_a$ in the far past. Its simple form and its ability to parametrize a wide range of theoretical DE models render it a particularly appealing choice. 
    \item i-JBP: The Jassal-Bagla-Padmanabhan (JBP) parametrization \citep{jbp}, another well-discussed DE parametrization that proposes the following form for the DE EoS:
    \begin{equation}
        w(a)=w_0+w_a \, a \, (1-a)\:\:.
    \end{equation}
    \item i-MEDE: The modified emergent DE (MEDE) model is a simple generalization of the phenomenological emergent DE (PEDE) model \citep{Li_2019}, obtained by including one additional degree of freedom $\alpha$. Here, the DE EoS takes the form \citep{Benaoum:2020qsi}
    \begin{equation}
        w(a)=-1-\dfrac{\alpha}{3\ln10}\left[1-\tanh\left(\alpha \log_{10}a\right)\right]\:\:.
    \end{equation}
    If $\alpha\,=\,0$, the model reduces to the $\Lambda$CDM scenario, while for $\alpha\,=\,1$ the PEDE model is recovered. In this work, we re-parametrize $\alpha$ as $w_0=-1-\frac{\alpha}{3\ln10}$.
\end{enumerate}
The following data sets are utilized in our analysis:
\begin{itemize}
    \item \textbf{CMB}: CMB temperature and polarization angular power spectra, and CMB lensing of \textit{Planck} 2018 \textit{TTTEEE+low l+low E+lensing} \citep{Pl18V,Pl18VI,Pl18VIII}.
    \item \textbf{BAO}: BAO measurements by 6 degree Field Galaxy Survey (6dFGS) \citep{6dF}, Sloan Digital Sky Survey Main Galaxy Sample (SDSS MGS) \citep{MGS}, BOSS Data Release 12 (DR12) (LOWZ, CMASS, and luminous red galaxy (LRG)) \citep{bao1}, and Extended Baryon Oscillation Spectroscopic Survey (eBOSS) DR16 (galaxies, LRG, quasi-stellar object (QSO), emission-line galaxy (ELG), Lyman-alpha (Ly$\alpha$), and Ly$\alpha\times$QSO) \citep{eBOSS:2020lta,eBOSS:2020hur,eBOSS:2020gbb,eBOSS:2020uxp,eBOSS:2020qek,eBOSS:2020tmo,eBOSS:2020yzd,eBOSS:2020fvk}. 
    \item \textbf{SNIa}: Luminosity distance data from 1701 light curves representing 1550 distinct SNIa from the Pantheon+ compilation \citep{panprelease}.
\end{itemize}
As argued in \citet{Shah:2023}, we deliberately leave Supernovae and H0 for the Equation of State (SH0ES) \citep{Riess:2021jrx} data out of this analysis in order to do an honest comparison of the $H_0$ values as obtained from the above data sets against the SH0ES value, so as to give a proper estimation of the Hubble tension (summarized in Table \ref{tab:tensions}).

\begin{table}
    \begin{center}
    \resizebox{0.5\textwidth}{!} {\renewcommand{\arraystretch}{1.1} \setlength{\tabcolsep}{25 pt}
        \begin{tabular}{c c}
            \hline
            \textbf{Parameter}             & \textbf{Prior}\\
            \hline
            $\Omega_{\rm b} h^2$           & $[0.005,   0.1]$\\
            $\Omega_{\rm c} h^2$           & $[0.01,    0.99]$\\
            $100\theta_{s}$                & $[0.5,     10]$\\
            $\ln\left(10^{10}A_{s}\right)$ & $[1,       5]$\\
            $n_{s}$                        & $[0.5,     1.5]$\\
            $\tau$                         & $[0,       0.9]$\\
            $Q$                            & $[-2,      2]$\\
            $w_0$                          & $[-1,      0]$ in non-phantom regime\\
            $w_0$                          & $[-3,      -1]$ in phantom regime\\
            $w_a$                          & $[-2,      2]$\\
            $c^2_{s}$                      & $[0,       2]$\\
            \hline            
        \end{tabular}
        }
        \caption{Priors used for obtaining constraints on the cosmological parameters of interest.}
        \label{tab:priors}
        \vspace{-11pt}
    \end{center}
\end{table}
We implement the models under consideration into the Boltzmann solver code \texttt{CLASS}\footnote{\url{https://github.com/lesgourg/class_public}} \citep{CLASS1,CLASS2}, building upon the publicly available modified versions by \citet{Hoerning:2023hks,Lucca:2020zjb}, by employing Romberg integration methods for evaluating certain otherwise intractable integrals involving the dynamical EoS of DE. To prevent any instabilities arising from the DE EoS $w(z)$ potentially crossing -1, crucial checks within \texttt{CLASS} ensure that $w(z)$ does not transition through -1 between the infinite past and the present day $z=0$. We also properly take into account the effect of the C/DDE parametrizations on the computation of the growth factor \citep{Linder:2003dr}. Constraints are obtained on the model parameters using \texttt{MontePython}\footnote{\url{https://github.com/brinckmann/montepython_public}} \citep{mcmc,MontePython2}, with the priors considered for the Markov chain Monte Carlo (MCMC) analyses as given in Table \ref{tab:priors}, by ensuring strict non-phantom and phantom bounds for the DE EoS parameters - $w_0$ and $w_a$. For this, we set a hard-bound $\left(w(z=0)+1\right) \cdot \left(w(z)+1\right) > 0$; this condition ensures that the DE model remains in a stable regime throughout its evolution and prevents the occurrence of unphysical perturbations that could arise from crossing the phantom divide. The corresponding contour plots were generated using \texttt{GetDist}\footnote{\url{https://github.com/cmbant/getdist}} \citep{Lewis:2019xzd}.

\section{Parameter estimation using MCMC}\label{sec:results}
We now present the results of our MCMC analyses on a case-by-case basis, with separate discussions on the phantom and non-phantom regimes for the DE sector. Such a treatment is justified due to various reasons. First, the presence of a ``doom factor'' [$1/(1+w)$] in the perturbation equations causes them to become singular if the EoS of the DE model undergoes phantom-crossing (at $w=-1$) at any point of cosmic evolution. Secondly, the inherently distinct nature of DE in the two regimes about the phantom divide line $w=-1$ necessitates a careful examination of each sector independently. Crossing the phantom divide line has two cosmological implications - either the DE consists of multiple components with at least one non-canonical phantom component, or general relativity needs to be extended to a more general theory on cosmological scales \citep{Nesseris:2006er, Hu:2004kh, Fang:2008sn}. Moreover, when subjected to observational data, the two regimes exhibit distinct qualitative constraints on other cosmological parameters (see, e.g. \citet{Hazra:2013dsx,Novosyadlyj:2013nya,Bhattacharyya:2018fwb}). These differences underscore the importance of treating them separately to ensure a robust theoretical framework for reliable and accurate interpretations of observational data.

\textbf{Note}: Throughout the paper, we fix the sound speed of DE as $c_s^2=1$, except when it is explicitly mentioned to be varying (Sec. \ref{cs2free}). Moreover, we refer to each model by its usual name when $Q=0$, and we prefix it with ``i-" when $Q$ is kept open as a free parameter.

\subsection{Non-phantom regime for DE}
\begin{table*}
    \resizebox{1.0\textwidth}{!}{\renewcommand{\arraystretch}{1.25} \setlength{\tabcolsep}{15 pt}
    \begin{tabular}{c c c c c c}
        \hline\hline
        \textbf{Parameters} & \textbf{i-}$\boldsymbol{\Lambda}$\textbf{CDM} & \textbf{i-$w$CDM} & \textbf{i-MEDE} & \textbf{i-CPL} & \textbf{i-JBP} \\ \hline
        {\boldmath${\Omega_b}{h^2}$} & $0.02244\pm 0.00014        $ & $0.02243\pm 0.00013        $ & $0.02242\pm 0.00014        $ & $0.02243\pm 0.00014        $ & $0.02240\pm 0.00014        $\\

        {\boldmath${\Omega_c}{h^2}$} & $0.1254^{+0.0078}_{-0.0070}$ & $0.094^{+0.033}_{-0.014}   $ & $0.066^{+0.028}_{-0.042}   $ & $0.071^{+0.035}_{-0.028}   $ & $0.105^{+0.042}_{-0.021}   $\\
        
        {\boldmath$100{\theta_s}$} & $1.04195\pm 0.00029        $ & $1.04196\pm 0.00029        $ & $1.04192\pm 0.00029        $ & $1.04194\pm 0.00029        $ & $1.04188\pm 0.00029        $\\
        
        {\boldmath${\ln{\left({10^{10}A_s}\right)}}$} & $3.048\pm 0.015            $ & $3.049\pm 0.015            $ & $3.046\pm 0.015            $ & $3.048\pm 0.014            $ & $3.044\pm 0.014            $\\
        
        {\boldmath$n_s$          } & $0.9672\pm 0.0037          $ & $0.9675\pm 0.0038          $ & $0.9672\pm 0.0037          $ & $0.9676\pm 0.0038          $ & $0.9660\pm 0.0040          $\\
        
        {\boldmath${\tau}$       } & $0.0564^{+0.0069}_{-0.0077}$ & $0.0566^{+0.0069}_{-0.0078}$ & $0.0551\pm 0.0075          $ & $0.0562\pm 0.0073          $ & $0.0541\pm 0.0074          $\\
        
        {\boldmath$Q$            } & $0.061\pm 0.078            $ & $-0.21^{+0.26}_{-0.15}     $ & $-0.42^{+0.16}_{-0.32}     $ & $-0.38^{+0.21}_{-0.25}     $ & $-0.11^{+0.34}_{-0.24}     $\\
        
        {\boldmath$w_0$          } &            -                  & $-0.907^{+0.030}_{-0.089}  $ & $-0.801^{+0.13}_{-0.087}   $ & $-0.829\pm 0.088           $ & $-0.811^{+0.089}_{-0.12}   $\\
        
        {\boldmath$w_a$          } &             -                 &              -                &              -                & $-0.09\pm 0.11             $ & $-0.90\pm 0.47             $\\
        
        \hline
        
        {\boldmath$H_0$          } & $67.09\pm 0.67             $ & $67.03\pm 0.68             $ & $67.07\pm 0.65             $ & $67.00\pm 0.67             $ & $67.15\pm 0.69             $\\
        
        {\boldmath$\Omega_{m0}$  } & $0.330\pm 0.023            $ & $0.260^{+0.075}_{-0.036}   $ & $0.198^{+0.072}_{-0.094}   $ & $0.211^{+0.077}_{-0.067}   $ & $0.284^{+0.093}_{-0.048}   $\\
        
        {\boldmath$\sigma_{8,0}$ } & $0.778^{+0.038}_{-0.050}   $ & $1.037^{+0.024}_{-0.30}    $ & $1.44^{+0.16}_{-0.66}      $ & $1.33^{+0.12}_{-0.56}      $ & $0.983^{+0.025}_{-0.32}    $\\
        
        {\boldmath$S_8$ } & $0.813_{-0.020}^{+0.023}$ & $0.880_{-0.054}^{+0.128}$ & $1.017_{-0.147}^{+0.261}$ & $0.976_{-0.111}^{+0.228}$ & $0.855_{-0.063}^{+0.144}$ \\
        
        \hline 
        {\boldmath $\chi^2_{min}$ } & 4203 & 4204 & 4204 & 4204 & 4201 \\
        {\boldmath $-\ln{\cal L}_\mathrm{min}$ } & 2101.69 & 2101.92 & 2101.77 & 2101.75 & 2100.52 \\
        \hline
        \hline
    \end{tabular}
    }
\caption{The mean and 1$\sigma$ constraints obtained for the models considered in section \ref{sec:models_datasets} using combined \textit{Planck} 2018 + BAO + Pantheon+ observational data, in the non-phantom regime for the i-C/DDE parametrizations.}
\label{tab:nonphantconstraintscs2one}
\vspace{-12pt}
\end{table*}

\begin{figure*}
    \centering
    \includegraphics[width=\textwidth]{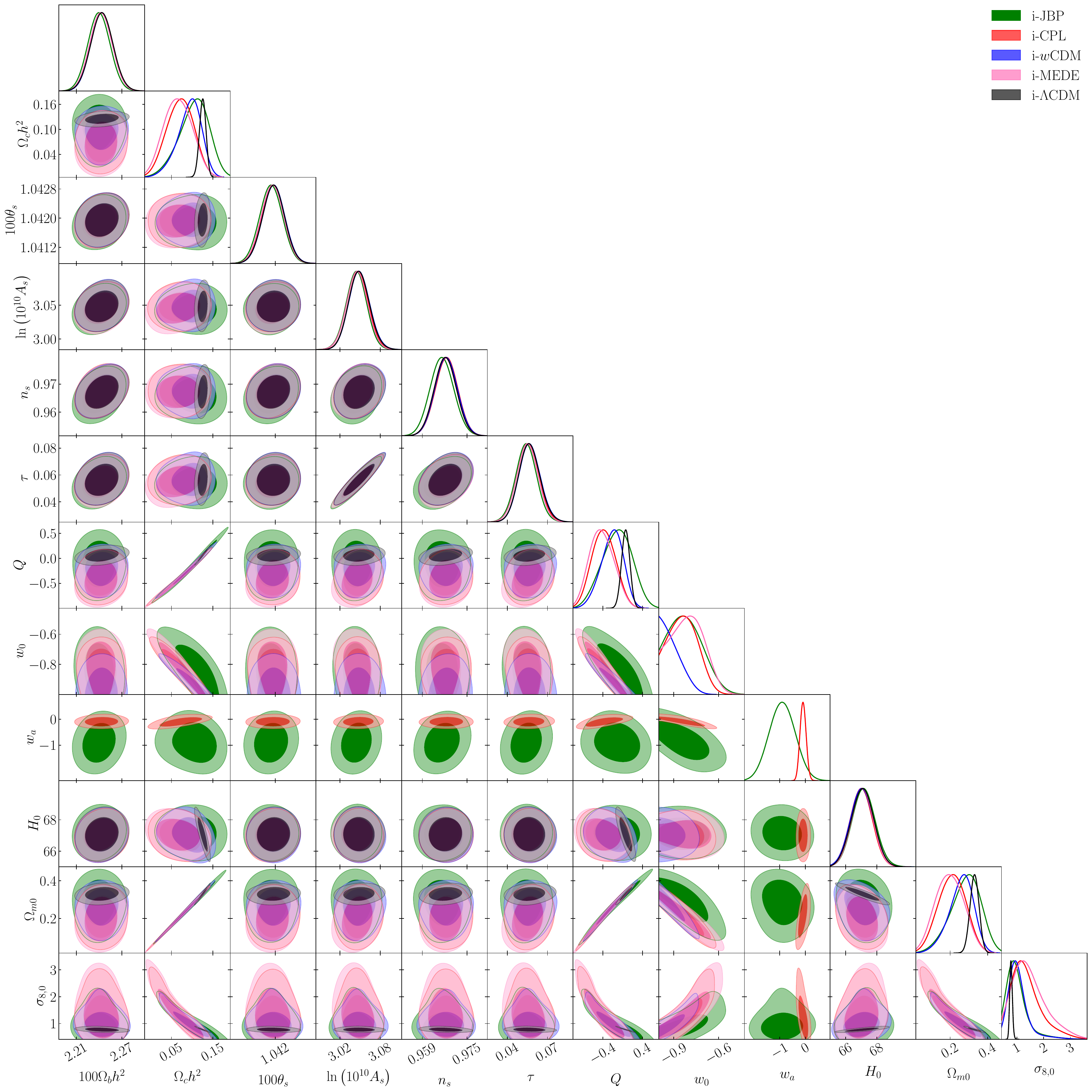}
    \caption{Comparison of constraints obtained for the models considered in section \ref{sec:models_datasets} using combined \textit{Planck} 2018 + BAO + Pantheon+ observational data, in the non-phantom regime for the i-C/DDE parametrizations ($c_s^2=1$).}
    \label{fig:nonphanttriangle}
\end{figure*}

\begin{table*}
    \resizebox{1.0\textwidth}{!}{\renewcommand{\arraystretch}{1.25} \setlength{\tabcolsep}{15 pt} 
    \begin{tabular}{c c c c c c}
        \hline\hline
        \textbf{Parameters} & \textbf{i-}$\boldsymbol{\Lambda}$\textbf{CDM} & \textbf{i-$w$CDM} & \textbf{i-MEDE} & \textbf{i-CPL} & \textbf{i-JBP} \\ \hline
        {\boldmath${\Omega_b}{h^2}$} & $0.02244\pm 0.00014        $ & $0.02243\pm 0.00014        $ & $0.02243\pm 0.00014        $ & $0.02240\pm 0.00014        $ & $0.02242\pm 0.00014        $\\

        {\boldmath${\Omega_c}{h^2}$} & $0.1254^{+0.0078}_{-0.0070}$ & $0.143\pm 0.012            $ & $0.143^{+0.014}_{-0.012}   $ & $0.151^{+0.012}_{-0.0067}  $ & $0.146^{+0.017}_{-0.0080}  $\\
        
        {\boldmath$100{\theta_s}$} & $1.04195\pm 0.00029        $ & $1.04194\pm 0.00029        $ & $1.04193\pm 0.00029        $ & $1.04189\pm 0.00029        $ & $1.04193\pm 0.00030        $\\
        
        {\boldmath${\ln{\left({10^{10}A_s}\right)}}$} & $3.048\pm 0.015            $ & $3.048\pm 0.014            $ & $3.047\pm 0.015            $ & $3.044\pm 0.014            $ & $3.047\pm 0.015            $\\
        
        {\boldmath$n_s$          } & $0.9672\pm 0.0037          $ & $0.9669\pm 0.0038          $ & $0.9668\pm 0.0038          $ & $0.9658\pm 0.0038          $ & $0.9666\pm 0.0038          $\\
        
        {\boldmath${\tau}$       } & $0.0564^{+0.0069}_{-0.0077}$ & $0.0561\pm 0.0074          $ & $0.0558^{+0.0068}_{-0.0077}$ & $0.0541\pm 0.0074          $ & $0.0556\pm 0.0074          $\\
        
        {\boldmath$Q$            } & $0.061\pm 0.078            $ & $0.25^{+0.13}_{-0.16}      $ & $0.25\pm 0.14              $ & $0.333^{+0.14}_{-0.087}    $ & $0.29^{+0.18}_{-0.11}      $\\
        
        {\boldmath$w_0$          } &            -                  & $-1.066^{+0.061}_{-0.025}  $ & $-1.064^{+0.059}_{-0.025}  $ & $-1.047^{+0.046}_{-0.012}  $ & $-1.061^{+0.061}_{-0.016}  $\\
        
        {\boldmath$w_a$          } &             -                 &             -                 &              -                & $-0.22^{+0.18}_{-0.15}     $ & $-0.14^{+0.36}_{-0.44}     $\\
        
        \hline
        
        {\boldmath$H_0$          } & $67.09\pm 0.67             $ & $67.18\pm 0.67             $ & $67.24\pm 0.68             $ & $67.37\pm 0.66             $ & $67.25\pm 0.68             $\\
        
        {\boldmath$\Omega_{m0}$  } & $0.330\pm 0.023            $ & $0.368\pm 0.030            $ & $0.368^{+0.034}_{-0.027}   $ & $0.384^{+0.028}_{-0.019}   $ & $0.375^{+0.037}_{-0.023}   $\\
        
        {\boldmath$\sigma_{8,0}$ } & $0.778^{+0.038}_{-0.050}   $ & $0.709^{+0.039}_{-0.061}   $ & $0.708^{+0.035}_{-0.062}   $ & $0.683^{+0.023}_{-0.045}   $ & $0.697^{+0.027}_{-0.062}   $\\
        
        {\boldmath$S_8$ } & $0.813_{-0.020}^{+0.023}$ & $0.780_{-0.024}^{+0.028}$ & $0.779_{-0.023}^{+0.028}$ & $0.769_{-0.017}^{+0.021}$ & $0.773_{-0.021}^{+0.028}$ \\

        \hline 
        {\boldmath $\chi^2_{min}$ } & 4203 & 4203 & 4202 & 4199 & 4201 \\
        {\boldmath $-\ln{\cal L}_\mathrm{min}$ } & 2101.69 & 2101.3 & 2100.75 & 2099.4 & 2100.5 \\
        \hline
        \hline
    \end{tabular}
    }
\caption{The mean and 1$\sigma$ constraints obtained for the models considered in section \ref{sec:models_datasets} using combined \textit{Planck} 2018 + BAO + Pantheon+ observational data, in the phantom regime for the different DE parametrizations.}
\label{tab:phantconstraintscs2one}
\vspace{-12pt}
\end{table*}

Table \ref{tab:nonphantconstraintscs2one} shows the constraints on the interacting models considered in the non-phantom regime. The full triangle plots are in Fig. \ref{fig:nonphanttriangle}. 

\begin{figure*} 
\begin{center}
    \begin{subfigure}{.32\textwidth}
        \includegraphics[width=\textwidth]{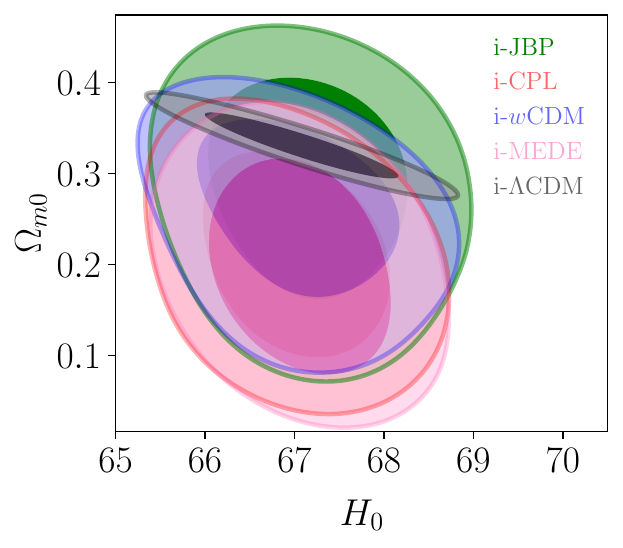}
    \end{subfigure}
    \begin{subfigure}{.32\textwidth}
        \includegraphics[width=\textwidth]{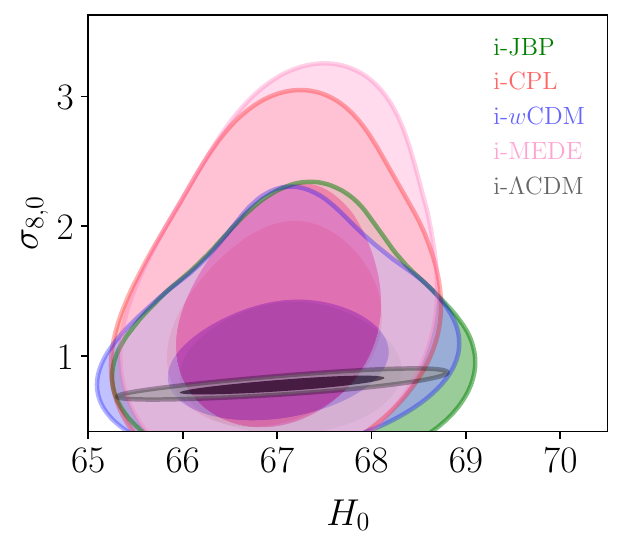}
    \end{subfigure}
    \begin{subfigure}{.32\textwidth}
        \includegraphics[width=\textwidth]{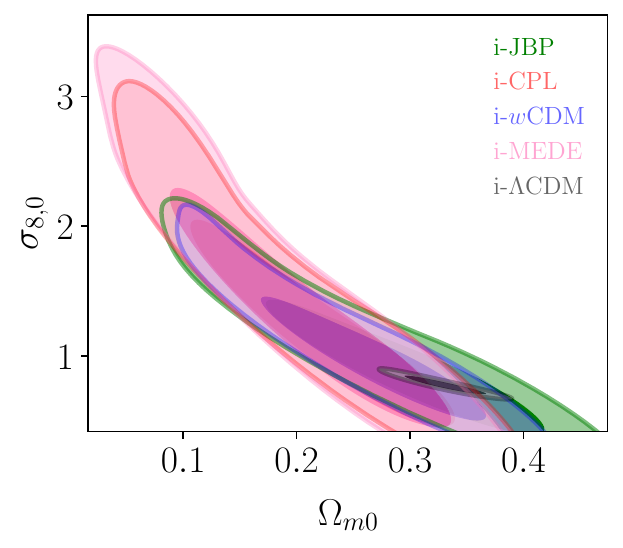}
    \end{subfigure}
\end{center}
    \vspace*{-0.5cm}
    \caption{Constraints on and correlations between $H_0$, $\Omega_{m0}$, and $\sigma_{8,0}$ in the non-phantom regime.}
    \label{fig:nonphantH0OmS8}
    \vspace*{-10pt}
\end{figure*}

Fig. \ref{fig:nonphantH0OmS8} highlights the major cosmological parameters of our concern. We note the following:
\begin{itemize}
    \item In comparison to i-$\Lambda$CDM, all the DE parametrizations exhibit a reduced level of correlation between $H_0$ and $\Omega_{m0}$.
    \item The strong correlation between $\sigma_{8,0}$ and $H_0$ is somewhat alleviated, with $\sigma_{8,0}$ preferring higher values than in i-$\Lambda$CDM. However, there is no shift in the values of $H_0$, with the best-fitting values remaining consistent with early-time estimations for i-$\Lambda$CDM.
    \item $\Omega_{m0}$ and $\sigma_{8,0}$ correlation is preserved. All i-C/DDE cases favour a lower $\Omega_{m0}$ value, and hence a higher value of $\sigma_{8,0}$.
\end{itemize}

\begin{figure*}
\begin{center}
    \begin{subfigure}{.32\textwidth}
        \includegraphics[width=\textwidth]{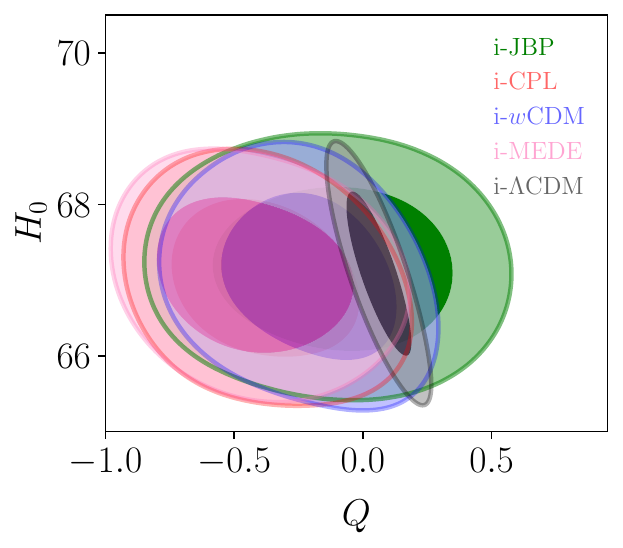}
    \end{subfigure}
    \begin{subfigure}{.32\textwidth}
        \includegraphics[width=\textwidth]{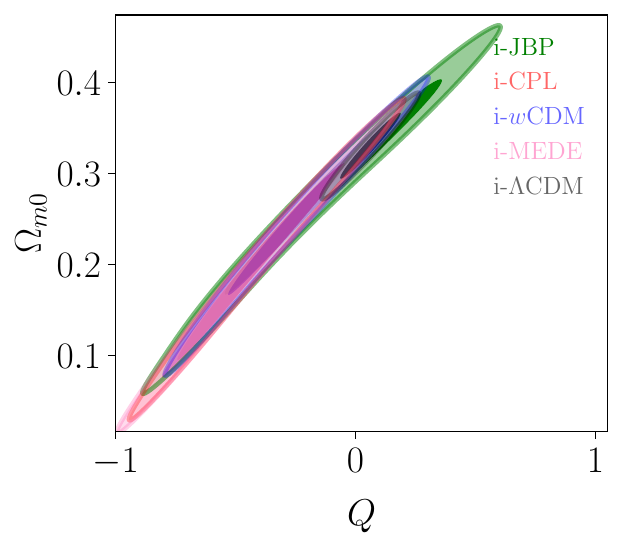}
    \end{subfigure}
    \begin{subfigure}{.32\textwidth}
        \includegraphics[width=\textwidth]{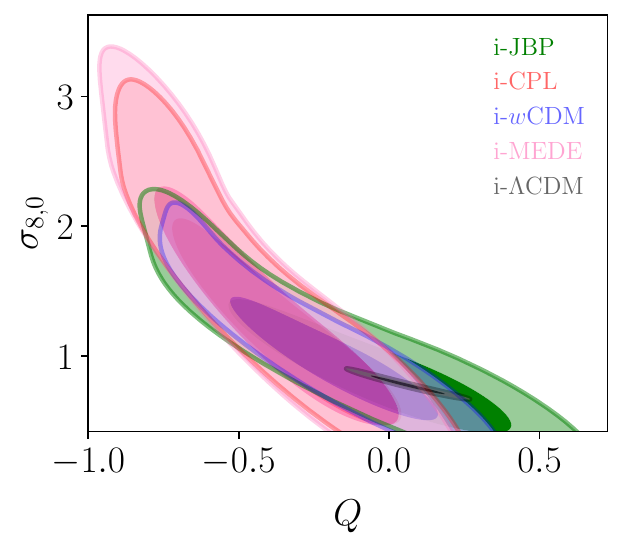}
    \end{subfigure}
\end{center}
    \vspace*{-0.5cm}
    \caption{Constraints on and correlations of $Q$ versus $H_0$, $\Omega_{m0}$, and $\sigma_{8,0}$ in the non-phantom regime.}
    \label{fig:nonphantQOmS8H0}
    \vspace*{-10pt}
\end{figure*}

Fig. \ref{fig:nonphantQOmS8H0} illustrates the effect of the interaction parameter. It reveals the following:
\begin{itemize}
    \item The correlation between $H_0$ and $Q$ is negligible for all models except i-$\Lambda$CDM.
    \item There is a strong positive correlation between $\Omega_{m0}$ and $Q$. All i-C/DDE models in the non-phantom regime prefer an energy flow from the DM to the DE sector.
    \item The parameters $\sigma_{8,0}$ and $Q$ exhibit a strong negative correlation, leading to a preference for negative $Q$ values for the i-C/DDE models, which in turn increases $\sigma_{8,0}$.
\end{itemize}

\begin{figure*} 
\begin{center}
    \begin{subfigure}{.32\textwidth}
        \includegraphics[width=\textwidth]{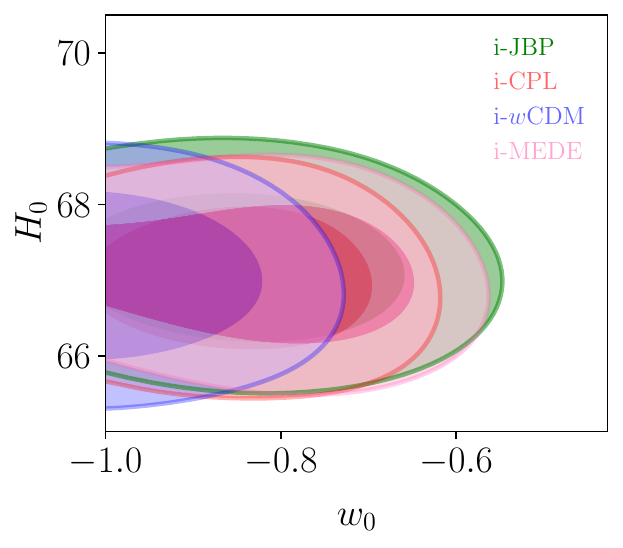}
    \end{subfigure}
    \begin{subfigure}{.32\textwidth}
        \includegraphics[width=\textwidth]{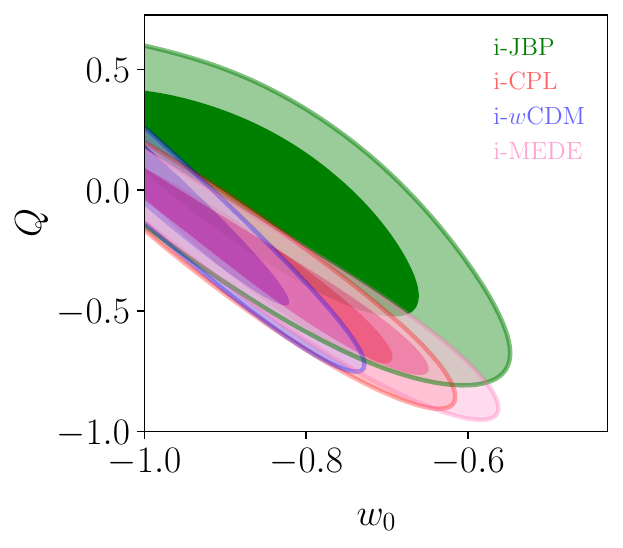}
    \end{subfigure}
    \begin{subfigure}{.32\textwidth}
        \includegraphics[width=\textwidth]{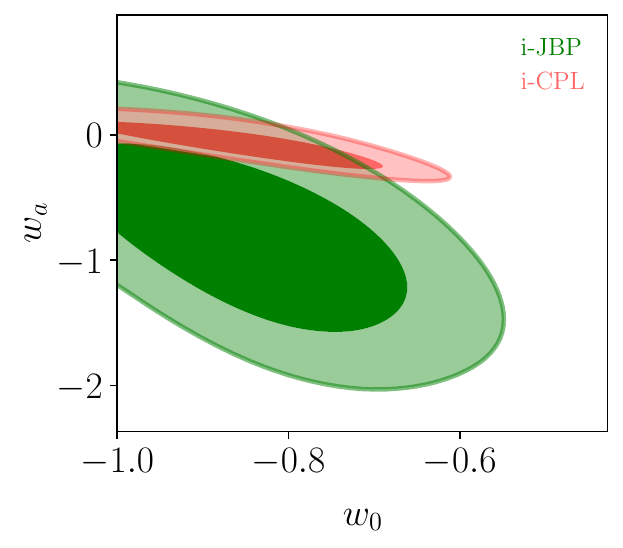}
    \end{subfigure}
\end{center}
    \vspace*{-0.5cm}
    \caption{Constraints on and correlations of $w_0$ versus $H_0$, $Q$ and $w_a$ in the non-phantom regime.}
    \label{fig:nonphantQH0w0wa}
    \vspace*{-10pt}
\end{figure*}

Finally, Fig. \ref{fig:nonphantQH0w0wa} depicts the effect of the EoS parameters. We find that
\begin{itemize}
    \item There is almost no correlation between $w_0$ and $H_0$, with $H_0$ centred around the early-time measurement.
    \item $w_0$ shows a strong negative correlation with $Q$, and hence a negative interaction prefers a more non-phantom nature.
    \item $w_0$ and $w_a$ are seen to be slightly negatively correlated.
\end{itemize}

\newpage

Based on these findings, we can infer that an interaction that allows energy transfer from the DM sector to the DE sector is favoured if the EoS is non-phantom. However, the most crucial point to note here is the value of $\sigma_{8,0}$ in Table \ref{tab:nonphantconstraintscs2one}. Generically, for any beyond - i-$\Lambda$CDM model, $\sigma_{8,0}$ takes a very large value, leading to overproduction of structures, which is naturally unrealistic and hence is not acceptable. Consequently, this tends to worsen the tension in $\sigma_{8,0}$ for all the i-C/DDE models, although the constraints on $H_0$ remain unaffected, indicating no improvement over the Hubble tension. This is mostly because of the very unrealistic value for the interaction term. So, although the presence of this interaction term helps in breaking the correlation between the parameters $H_0$ and $\sigma_{8,0}$, the non-phantom EoS fails miserably to address the $\sigma_{8,0}$ tension in the interacting scenario, generically for all the DE models here.

\subsection{Phantom regime for DE}
\begin{figure*} 
    \centering
    \includegraphics[width=\textwidth]{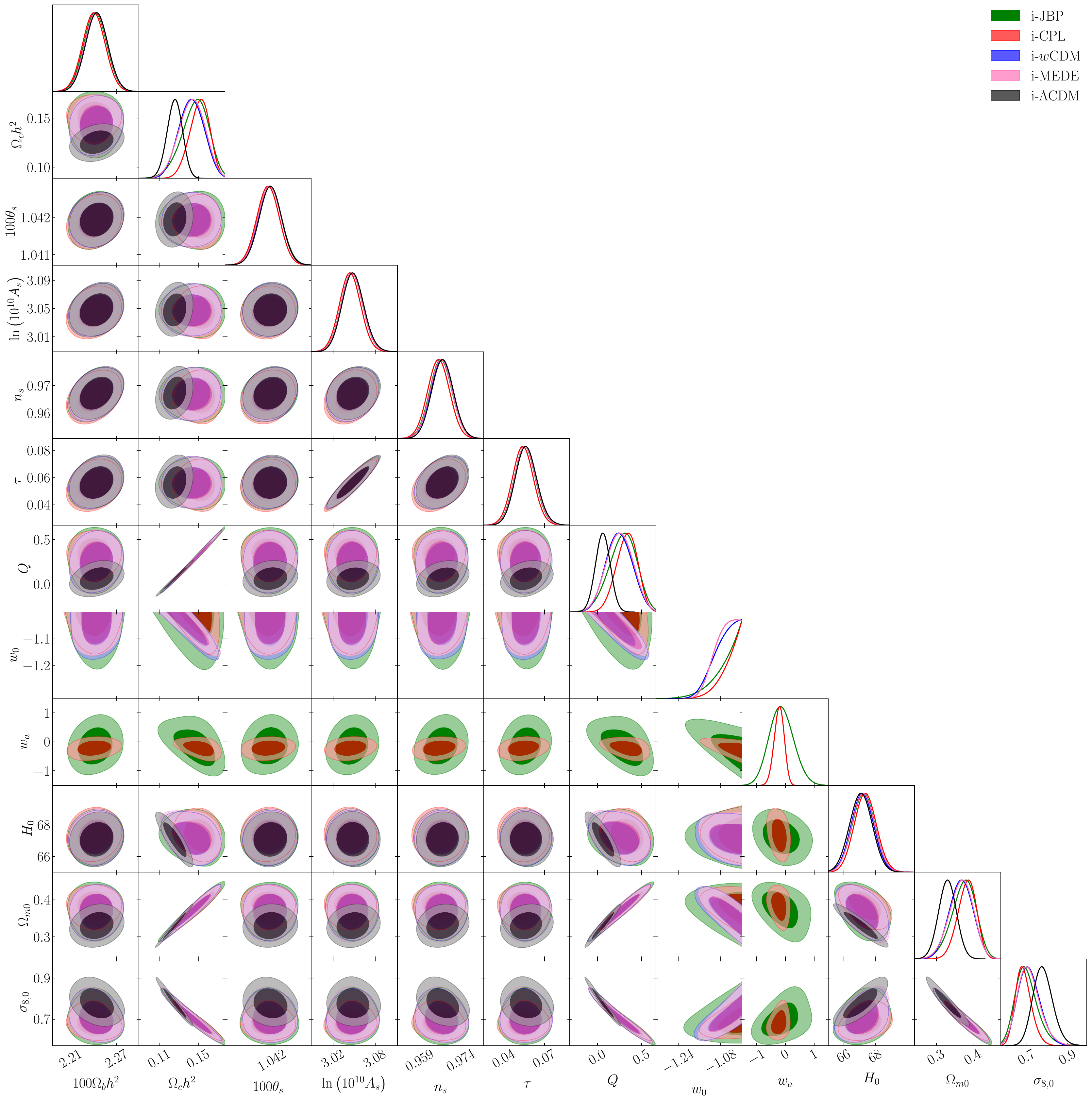}
    \caption{Comparison of constraints obtained for the models considered in section \ref{sec:models_datasets} using combined \textit{Planck} 2018 + BAO + Pantheon+ observational data, in the phantom regime for the i-C/DDE parametrizations ($c_s^2=1$).}
    \label{fig:phanttriangle}
    \vspace{-10pt}
\end{figure*}

Table \ref{tab:phantconstraintscs2one} presents the constraints on the interacting models considered in the phantom regime. For detailed visualizations, refer to the full triangle plots in Fig. \ref{fig:phanttriangle}.

\begin{figure*} 
\begin{center}
    \begin{subfigure}{.32\textwidth}
        \includegraphics[width=\textwidth]{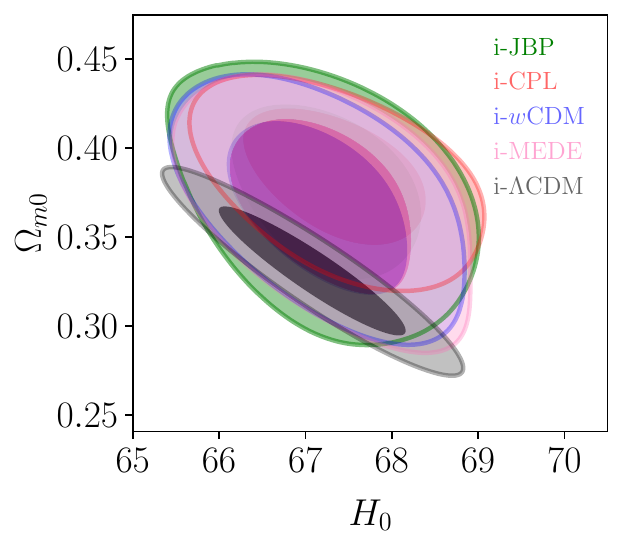}
    \end{subfigure}
    \begin{subfigure}{.32\textwidth}
        \includegraphics[width=\textwidth]{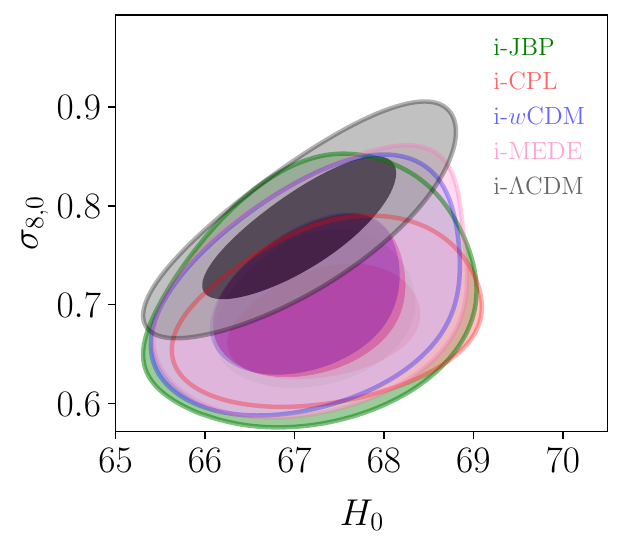}
    \end{subfigure}
    \begin{subfigure}{.32\textwidth}
        \includegraphics[width=\textwidth]{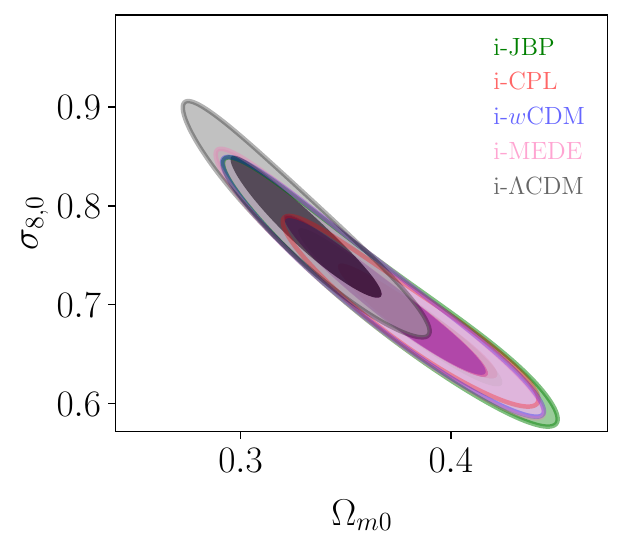}
    \end{subfigure}
\end{center}
    \vspace*{-0.5cm}
    \caption{Constraints on and correlations between $H_0$, $\Omega_{m0}$, and $\sigma_{8,0}$ in the phantom regime.}
    \label{fig:phantH0OmS8}
\end{figure*}

\begin{figure*} 
\begin{center}
    \begin{subfigure}{.32\textwidth}
        \includegraphics[width=\textwidth]{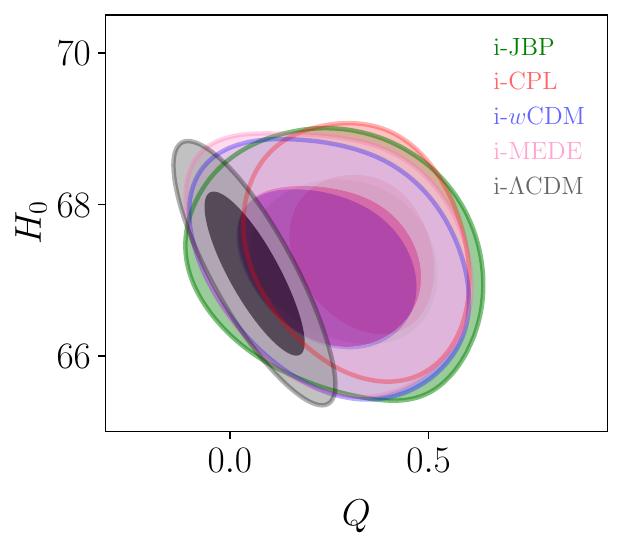}
    \end{subfigure}
    \begin{subfigure}{.32\textwidth}
        \includegraphics[width=\textwidth]{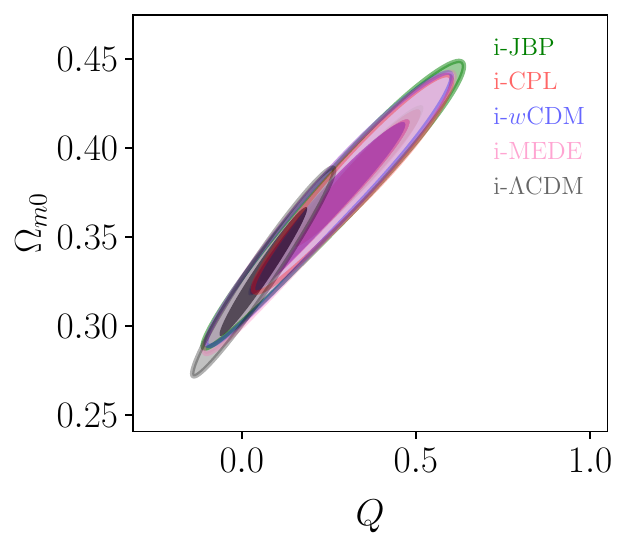}
    \end{subfigure}
    \begin{subfigure}{.32\textwidth}
        \includegraphics[width=\textwidth]{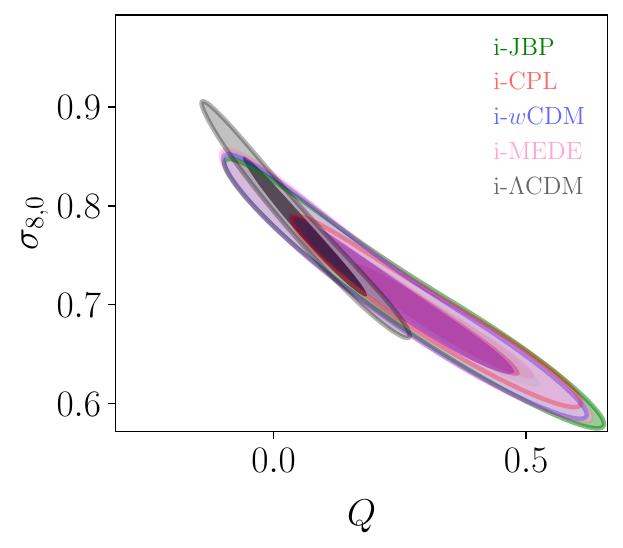}
    \end{subfigure}
\end{center}
    \vspace*{-0.5cm}
    \caption{Constraints on and correlations of $Q$ versus $H_0$, $\Omega_{m0}$, and $\sigma_{8,0}$ in the phantom regime.}
    \label{fig:phantQOmS8H0}
\end{figure*}

\begin{figure*} 
\begin{center}
    \begin{subfigure}{.32\textwidth}
        \includegraphics[width=\textwidth]{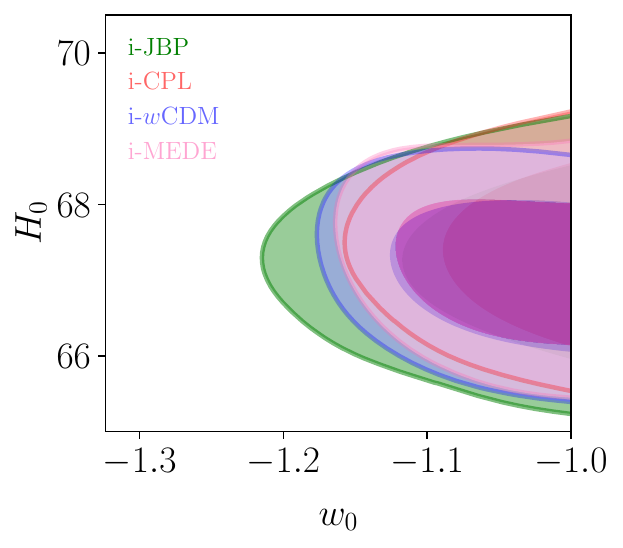}
    \end{subfigure}
    \begin{subfigure}{.32\textwidth}
        \includegraphics[width=\textwidth]{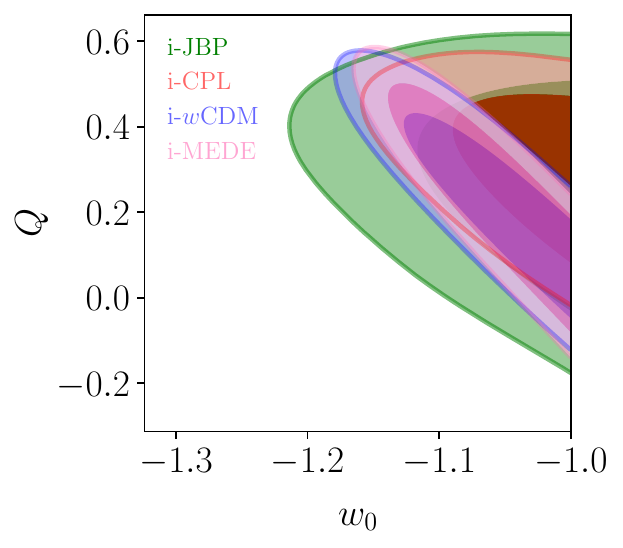}
    \end{subfigure}
    \begin{subfigure}{.32\textwidth}
        \includegraphics[width=\textwidth]{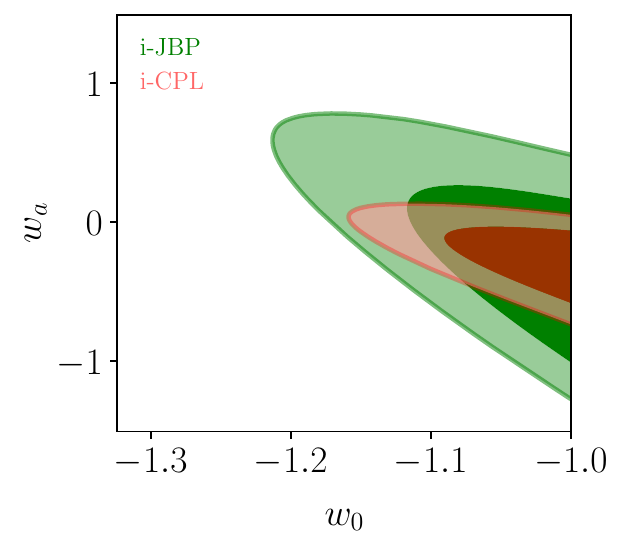}
    \end{subfigure}
\end{center}
    \vspace*{-0.5cm}
    \caption{Constraints on and correlations of $w_0$ versus $H_0$, $Q$, and $w_a$ in the phantom regime.}
    \label{fig:phantQH0w0wa}
\end{figure*}

As in the previous section, we separately highlight some of the parameters in Fig. \ref{fig:phantH0OmS8}, \ref{fig:phantQOmS8H0}, and \ref{fig:phantQH0w0wa}, in order to help us in the analysis. In Fig. \ref{fig:phantH0OmS8}, we observe the following trends:
\begin{itemize}
    \item All the DE parametrizations exhibit reduced correlation between $H_0$ and $\Omega_{m0}$ compared to i-$\Lambda$CDM.
    \item The strong positive correlation between $H_0$ and $\sigma_{8,0}$ is relaxed in all the models in comparison to i-$\Lambda$CDM. This accommodates lower values of $\sigma_{8,0}$ without any considerable increase in $H_0$.
    \item The correlation between $\Omega_{m0}$ and $\sigma_{8,0}$ is preserved, with all i-C/DDE cases favouring higher values of $\Omega_{m0}$ and consequently lower values of $\sigma_{8,0}$.
\end{itemize}

Fig. \ref{fig:phantQOmS8H0} illustrates the following:
\begin{itemize}
    \item The correlation between $H_0$ and $Q$ is eliminated for all the interacting models except i-$\Lambda$CDM.
    \item There is a strong positive correlation between $\Omega_{m0}$ and $Q$. Like i-$\Lambda$CDM, all i-C/DDE models in the phantom regime indicate an energy flow from DE to DM.
    \item $\sigma_{8,0}$ and $Q$ exhibit a strong negative correlation. Therefore, a preference for a positive $Q$ contributes to a reduction in $\sigma_{8,0}$.\\\\
\end{itemize}

In Fig. \ref{fig:phantQH0w0wa} we find that
\begin{itemize}
    \item There exists almost zero correlation between $w_0$ and $H_0$, with $H_0$ centred around the early-time measurement.
    \item $w_0$ exhibits a negative correlation with $Q$, indicating that a positive interaction favours a more phantom-like behaviour. However, $w_a$ does not show any significant correlation with $Q$.
    \item The DE EoS parameters, $w_0$ and $w_a$, are negatively correlated.
\end{itemize}

Based on these observations, we find that an interaction that promotes energy transfer from the DE sector to the DM sector is favoured in the phantom scenario. This in turn helps lower the constraints on $\sigma_{8,0}$, thereby leading to chances of alleviating the clustering tension. Nevertheless, it does not lead to any unrealistic value for any of the parameters under consideration. It is important to note that the constraints on $H_0$, although still in tension with the late-time measurements, show the possibility of slight improvement (unlike the substantial improvement in $\sigma_{8,0}$) with simultaneous lowering of clustering tension. 

\subsection{Comparison and interpretation}
Let us now engage ourselves in a comparison of the outcome of phantom versus non-phantom regimes for the models under consideration. Our analysis reveals that an interacting scenario with a phantom DE sector helps in relaxing the clustering tension, with a slight improvement of the Hubble tension. This effect is achieved through a joint relaxation of the correlations between $H_0-\Omega_{m0}$ and $H_0-\sigma_{8,0}$, enabling $\Omega_{m0}$ and $\sigma_{8,0}$ to vary without inducing significant shifts in $H_0$. Interestingly, $H_0$ appears to lose its correlation to the interaction parameter $Q$ in the presence of the diverse EoS parametrization of DE. Strong correlations notably influence the behaviour of $\sigma_{8,0}$, while all DE models exhibit a negative correlation between $w_0$ and $Q$, indicating that a more phantom nature prefers higher values of $Q$. Given the existing degeneracy between $\Omega_{m0}$ and $\sigma_{8,0}$, this causes an increase in $\Omega_{m0}$ and a decrease in $\sigma_{8_0}$ due to their positive and negative correlations with $Q$, respectively. Such trends are also reflected in the positive correlation between $\Omega_{m0}$ and $\sigma_{8,0}$.

All correlations between the parameters are identically maintained for the non-phantom cases. However, the key difference arises due to the non-phantom nature of EoS, wherein $w_0$ is correlated with increasingly negative values of $Q$ as the model diverges further from a cosmological constant. This, in turn, leads to a decrease in $\Omega_{m,0}$ and an increase in $\sigma_{8,0}$, exacerbating the clustering tension. None the less, $H_0$ remains quite largely unaffected for the non-phantom scenario.

A closer examination of the observed correlations reveals that a positive $Q$ implies an injection of energy into the matter sector, thereby increasing the value of $\Omega_{m0}$. As $\Omega_{m0}$ and $\sigma_{8,0}$ are strongly degenerate, an increase in the value of $\Omega_{m0}$ is complemented with a decrease in $\sigma_{8,0}$. This is reflected through a negative correlation between $\sigma_{8,0}$ and $Q$. The inclusion of the Pantheon+ SNIa data puts tight constraints on the parameter space $w_0-w_a$ of the i-CPL and i-JBP models, with the presence of a strong negative correlation between $w_0$ and $w_a$ for both the phantom and non-phantom regimes. In all non-$\Lambda$CDM interacting DE models, there exists a negative correlation between $\Omega_{m0}$ and $w_0$, with an almost negligible correlation between $\Omega_{m0}$ and $w_a$, similar to non-interacting models. This observation strongly supports the negative correlation between $Q$ and $w_0$. Consequently, $H_0$ becomes decoupled from other cosmological parameters, such as $\Omega_{m0}$ and $w_0$, remaining mostly unaffected by shifts in these parameters in the presence of interacting scenarios for both the phantom and non-phantom regimes. 

\begin{table*}
    \resizebox{1.0\textwidth}{!}{\renewcommand{\arraystretch}{1.0} \setlength{\tabcolsep}{25 pt} 
    \begin{tabular}{c c c c c c}
        \hline\hline
        \textbf{Parameters} & \textbf{i-}$\boldsymbol{\Lambda}$\textbf{CDM} & \textbf{i-$w$CDM} & \textbf{i-MEDE} & \textbf{i-CPL} & \textbf{i-JBP} \\ 
        \hline
       \multirow{3}{0pt}{\boldmath$S_8$ } & $0.813_{-0.020}^{+0.023}$ & $0.780_{-0.024}^{+0.028}$ & $0.779_{-0.023}^{+0.028}$ & $0.769_{-0.017}^{+0.021}$ & $0.773_{-0.021}^{+0.028}$ \\
        & ($ 1.68 \sigma$) & ($ 0.60 \sigma$) & ($ 0.59 \sigma$) & ($ 0.33 \sigma$) & ($ 0.43 \sigma$) \\
        \hline
        \multirow{2}{0pt}{\boldmath$H_0$} & $67.090_{-0.677}^{+0.668}$ & $67.181_{-0.663}^{+0.668}$ & $67.226_{-0.658}^{+0.687}$ & $67.365_{-0.650}^{+0.659}$ & $67.256_{-0.684}^{+0.660}$ \\
        & ($ 4.79 \sigma$) & ($ 4.75 \sigma$) & ($ 4.72 \sigma$) & ($ 4.63 \sigma$) & ($ 4.65 \sigma$) \\
        \hline
        \hline
    \end{tabular}
    }
\caption{Indicative Gaussian tension metric values for the various models under study in the phantom regime. The tensions are computed against the late-time $H_0$ measurement by SH0ES \citep{Riess:2021jrx} ($H_0=73.04\pm1.04$) and the late-time $S_8/\sigma_{8,0}$ measurement from DES-Y3 \citep{DES:2021bvc} ($S_8=0.759^{+0.025}_{-0.023}/\sigma_{8,0}=0.783^{+0.073}_{-0.092}$), which are themselves in tension with early-time measurements assuming vanilla $\Lambda$CDM ($H_0=67.744_{-0.419}^{+0.406}, \sigma_{8,0}=0.810_{-0.006}^{+0.006}, S_8=0.824_{-0.010}^{+0.010}$), at $\sim5\sigma$ and $\sim3\sigma$, respectively.}
\label{tab:tensions}
\end{table*}

We find that a non-interacting scenario ($Q=0$) is almost always included at a 2$\sigma$ confidence level, irrespective of the phantom or non-phantom nature of DE. However, there are a few exceptions: i-MEDE in the non-phantom regime, and i-JBP in the phantom regime excludes $Q=0$ at a 2$\sigma$ confidence level. i-CPL shows a non-interacting scenario to be excluded at 3$\sigma$. These exceptions present interesting cases which warrant further investigation, which we aim to perform in a future work. These DE parametrizations, in phantom versus non-phantom limit, propose different scenarios for the interaction between the dark sectors, leading to different predictions for the direction of energy flow. In our work, a positive $Q$ indicates a transfer of energy from the DE to the DM sector implying that DE decays or converts into DM over cosmic time, whereas a negative $Q$ indicates the reverse, \textit{i.e.} an energy flow from the DM to the DE sector. We find that a phantom DE EoS favours an energy flow from the DE to the DM sector, whereas a non-phantom EoS for DE predicts a transfer of energy from the DM to the DE sector.

For the data sets under consideration, we observe a slight preference for phantom models over their non-phantom counterparts, as indicated by the $\chi^2$ and $-\ln{\cal L}_\mathrm{min}$ values. Although this improvement in $\chi^2$ values for phantom models is minor and should not be taken too seriously, it suggests a slightly better fit to observational data than non-phantom DE cases. While the non-phantom scenario indicates unrealistically high values for $\sigma_{8,0}$, which will eventually jeopardize structure formation by overproduction, phantom DE models lower the value of $\sigma_{8,0}$ that helps relax the $S_8$ tension without disturbing the other parameters significantly. Furthermore, phantom DE models not only demonstrate an improvement in the $S_8$ parameter but also do not worsen the Hubble tension. This further strengthens the case for phantom DE. These results align with previous findings \citep{Bhattacharyya:2018fwb} that a phantom EoS for DE, with energy flow from DE to DM, is slightly favoured based on current data. Overall, our analysis suggests that phantom DE models provide a more accurate description of the observed Universe and are marginally preferred over non-phantom models for the current combination of data sets.

Table \ref{tab:tensions} summarizes the extent of tension for both $S_8$ and $H_0$ for all the models under consideration where we quote the tension values against the late time $H_0$ measurement by SH0ES \citep{Riess:2021jrx} and $S_8$ measurement from the DES-Y3 \citep{DES:2021bvc} observations. We choose the $S_8/\sigma_{8,0}$ measurement from DES-Y3 as our reference point for quoting tension values. However, it is important to note that the precise constraints from weak lensing surveys may be cosmological model dependent. Therefore, the emphasis should be placed on the general trends in the relaxation of tensions rather than on the specific $\sigma$ values. It can be found that an interacting scenario with the cosmological constant is somewhat helpful with the $S_8$ tension. However, C/DDE interacting scenarios greatly help in relaxing the $S_8$ tension to $\lesssim1\sigma$, with i-CPL performing the best followed closely by i-JBP, and then i-MEDE and i-$w$CDM. Moreover, we note that a parametrization with two free parameters performs better than one free parameter only. For the Hubble tension, we see all the interacting scenarios to perform equally, and maintain the same to $\sim4.7\sigma$. Although this is not a fully satisfactory outcome, it is quite intriguing that these models (with the exception of interacting $\Lambda$CDM) tend to alleviate the $S_8$ tension this well, and also maintain the $H_0$ tension, rather than exacerbating it. This is thanks to the reduction, and in some cases complete nullification, of the correlation of $H_0$ with $\sigma_{8,0}$ and $\Omega_{m,0}$.

\subsection{Any effect of keeping \texorpdfstring{$c_s^2$}{} free?} \label{cs2free}
\begin{figure*}
    \centering
    \includegraphics[width=\textwidth]{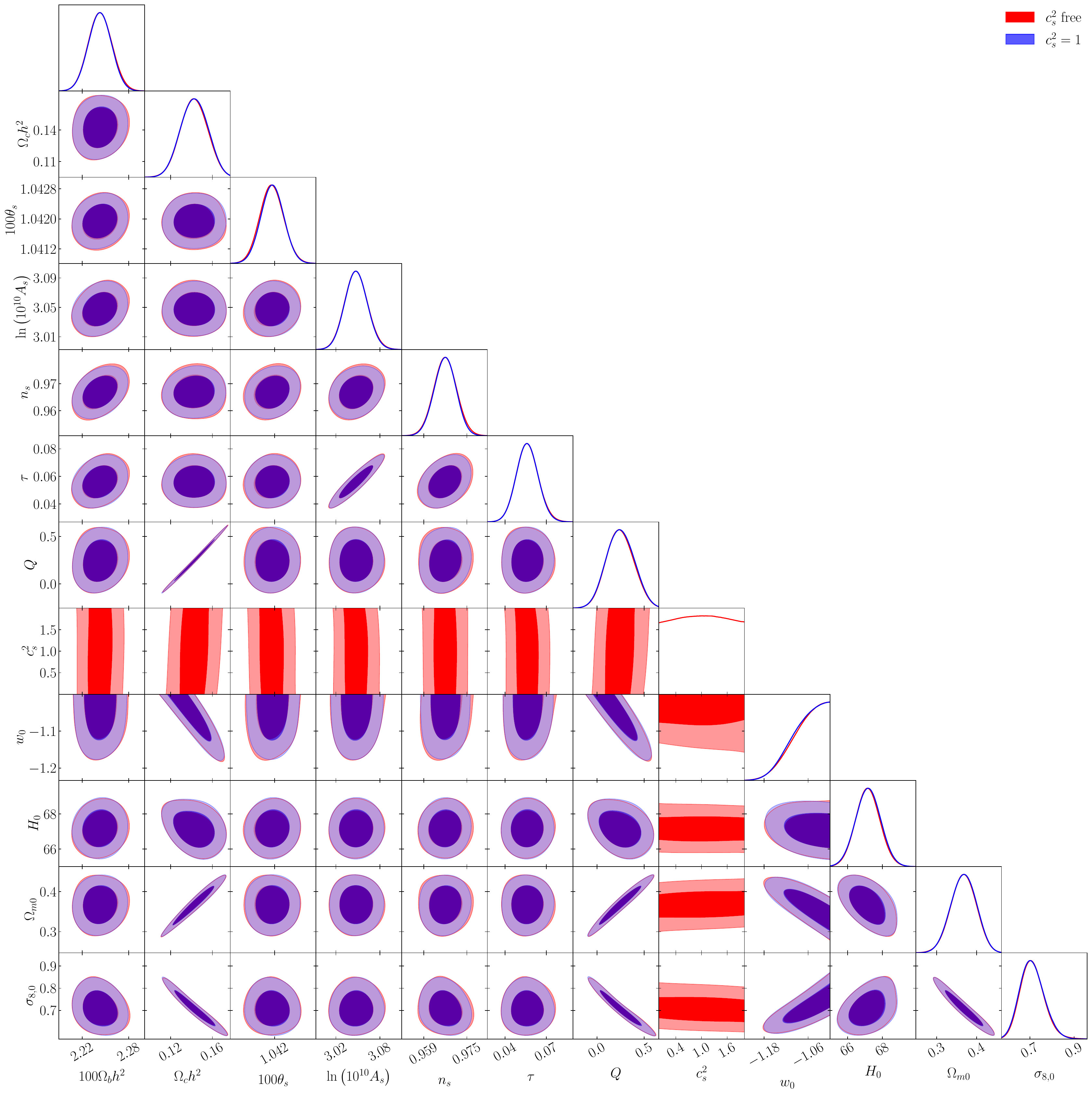}
    \caption{Comparison of constraints obtained for the i-$w$CDM model using combined \textit{Planck} 2018 + BAO + Pantheon+ observational data, in the phantom regime, once by fixing $c_s^2=1$, and once by keeping it as a free parameter in the MCMC analysis.}
    \label{fig:cs2compare}
    \vspace{-11pt}
\end{figure*}

\begin{table} 
    \resizebox{0.5\textwidth}{!}{\renewcommand{\arraystretch}{1.3} \setlength{\tabcolsep}{20 pt} 
    \begin{tabular}{c c c c}
        \hline\hline
        \textbf{Parameters} & $\boldsymbol{c_s^2=1}$ & $\boldsymbol{c_s^2}$ \textbf{free} \\ \hline
        {\boldmath${\Omega_b}{h^2}$} & $0.02243\pm 0.00014        $ & $0.02243\pm 0.00014        $\\

        {\boldmath${\Omega_c}{h^2}$} & $0.143\pm 0.012            $ & $0.142\pm 0.012            $\\
        
        {\boldmath$100{\theta_s}$} & $1.04194\pm 0.00029        $ & $1.04193\pm 0.00030        $\\
        
        {\boldmath${\ln{\left({10^{10}A_s}\right)}}$} & $3.048\pm 0.014            $ & $3.048\pm 0.015            $\\
        
        {\boldmath$n_s$          } & $0.9669\pm 0.0038          $ & $0.9670\pm 0.0040          $\\
        
        {\boldmath${\tau}$       } & $0.0561\pm 0.0074          $ & $0.0562^{+0.0069}_{-0.0079}$\\
        
        {\boldmath$Q$            } & $0.25^{+0.13}_{-0.16}      $ & $0.24^{+0.13}_{-0.15}      $\\
        
        {\boldmath$c_s^2$        } &             -                &  Unconstrained              \\
        
        {\boldmath$w_0$          } & $-1.066^{+0.061}_{-0.025}  $ & $-1.065^{+0.058}_{-0.025}  $\\
        
        \hline
        
        {\boldmath$H_0$          } & $67.18\pm 0.67             $ & $67.16\pm 0.65             $\\
        
        {\boldmath$\Omega_{m0}$  } & $0.368\pm 0.030            $ & $0.367\pm 0.030            $\\
        
        {\boldmath$\sigma_{8,0}$ } & $0.709^{+0.039}_{-0.061}   $ & $0.710^{+0.039}_{-0.058}   $\\
        
        {\boldmath$S_8$ } & $0.780_{-0.024}^{+0.028}$ & $0.781_{-0.024}^{+0.026}$ \\
        \hline
        \hline
    \end{tabular}
    }
\caption{The mean and 1$\sigma$ constraints obtained for the i-$w$CDM model using combined \textit{Planck} 2018 + BAO + Pantheon+ observational data, in the phantom regime, once by fixing $c_s^2=1$, and once by keeping it as a free parameter in the MCMC analysis.}
\label{tab:cs2freevsone}
\vspace{-1pt}
\end{table}

The sound speed of DE can influence the evolution of cosmological perturbations and, consequently, observables such as the CMB power spectrum or large-scale structure (LSS) observations. In order to examine any effect of that in this analysis, we run the code afresh with the same models and same data sets, this time keeping the DE sound speed free. We find that $c_s^2$ remains largely unconstrained; \textit{i.e.} it cannot be precisely probed by the given combination of data sets. The behaviour of DE for the above models is fairly consistent for the given range of $c_s^2$ values, indicating its degeneracy with other cosmological parameters. The lack of strong constraints on the sound speed also implies that the overall predictions of these iDMDE models are not overly sensitive to the specific value of $c_s^2$ in the considered range. Therefore, we find that for all the cases under study, keeping $c_s^2$ as a free parameter does not help in relaxing any tensions, does not affect any correlations significantly nor are there significant variations in the constraints of all the other parameters.  As a demonstrative example, we show in Fig. \ref{fig:cs2compare} and Table \ref{tab:cs2freevsone} the comparison between the constraints for the i-$w$CDM phantom interacting case, for $c_s^2=1$ and $c_s^2$ open, highlighting these features. 

\section{Possible role of \texorpdfstring{$\MakeLowercase{f}\sigma_8$}{} RSD data}\label{sec:RSD}
\begin{table*}
    \resizebox{1.0\textwidth}{!}{\renewcommand{\arraystretch}{1.25} \setlength{\tabcolsep}{15 pt} 
    \begin{tabular}{c c c c c c}
        \hline\hline
        \textbf{Parameters} & \textbf{i-}$\boldsymbol{\Lambda}$\textbf{CDM} & \textbf{i-$w$CDM} & \textbf{i-MEDE} & \textbf{i-CPL} & \textbf{i-JBP} \\ \hline
        {\boldmath${\Omega_b}{h^2}$} & $0.02243\pm 0.00013        $ & $0.02240\pm 0.00013        $ & $0.02240\pm 0.00014        $ & $0.02237\pm 0.00013        $ & $0.02242\pm 0.00014        $\\
        
        {\boldmath${\Omega_c}{h^2}$} & $0.1248\pm 0.0019          $ & $0.1255\pm 0.0019          $ & $0.1254\pm 0.0019          $ & $0.1261\pm 0.0019          $ & $0.1250\pm 0.0021          $\\
        
        {\boldmath$100{\theta_s}$} & $1.04195\pm 0.00028        $ & $1.04191\pm 0.00028        $ & $1.04190\pm 0.00029        $ & $1.04187\pm 0.00028        $ & $1.04193\pm 0.00029        $\\
        
        {\boldmath${\ln{\left({10^{10}A_s}\right)}}$} & $3.048\pm 0.014            $ & $3.046\pm 0.014            $ & $3.046\pm 0.015            $ & $3.043\pm 0.014            $ & $3.047\pm 0.015            $\\
        
        {\boldmath$n_s$          } & $0.9669\pm 0.0035          $ & $0.9661\pm 0.0037          $ & $0.9660\pm 0.0040          $ & $0.9651\pm 0.0036          $ & $0.9666\pm 0.0039          $\\
        
        {\boldmath${\tau}$       } & $0.0562\pm 0.0072          $ & $0.0550\pm 0.0073          $ & $0.0547\pm 0.0079          $ & $0.0533\pm 0.0071          $ & $0.0557\pm 0.0075          $\\
        
        {\boldmath$Q$            } & $0.052^{+0.015}_{-0.016}   $ & $0.056^{+0.015}_{-0.016}   $ & $0.055\pm 0.017            $ & $0.058\pm 0.016            $ & $0.053\pm 0.016            $\\
        
        {\boldmath$w_0$          } &              -                & $-1.019^{+0.019}_{-0.0051} $ & $-1.020^{+0.019}_{-0.0061} $ & $-1.017^{+0.017}_{-0.0056} $ & $-1.039^{+0.039}_{-0.0086} $\\
        
        {\boldmath$w_a$          } &              -                &              -                &              -                & $-0.071^{+0.085}_{-0.041}  $ & $0.19^{+0.19}_{-0.29}      $\\
        
        \hline
        
        {\boldmath$H_0$          } & $67.14\pm 0.43             $ & $67.51^{+0.48}_{-0.54}     $ & $67.43^{+0.59}_{-0.43}     $ & $67.81^{+0.51}_{-0.57}     $ & $67.30\pm 0.64             $\\
        
        {\boldmath$\Omega_{m0}$  } & $0.3280\pm 0.0078          $ & $0.3259\pm 0.0078          $ & $0.3255\pm 0.0083          $ & $0.3244\pm 0.0078          $ & $0.3270\pm 0.0080          $\\
        
        {\boldmath$\sigma_{8,0}$ } & $0.7801\pm 0.0093          $ & $0.7837\pm 0.0096          $ & $0.784\pm 0.010            $ & $0.7872\pm 0.0099          $ & $0.781\pm 0.011            $\\
        
        {\boldmath$S_8$ } & $0.816_{-0.009}^{+0.010}$ & $0.817_{-0.010}^{+0.010}$ & $0.817_{-0.010}^{+0.009}$ & $0.818_{-0.010}^{+0.009}$ & $0.816_{-0.010}^{+0.010}$ \\
        
        \hline
        {\boldmath $\chi^2_{min}$ } & 4216 & 4216 & 4214 & 4215 & 4216 \\
        {\boldmath $-\ln{\cal L}_\mathrm{min}$ } & 2107.84 & 2107.81 & 2107.13 & 2107.39 & 2107.86 \\
        \hline
        \hline
    \end{tabular}
    }
\caption{The mean and 1$\sigma$ constraints obtained for all the models in the phantom regime, for $c_s^2=1$, using combined \textit{Planck} 2018 + BAO + Pantheon+ observational data in addition to $f\sigma_8$ RSD data (compilation taken from \citet{Sinha:2021tnr}).}
\label{tab:fsigma8}
\end{table*}

\begin{figure} 
    \centering
    \includegraphics[width=0.5\textwidth]{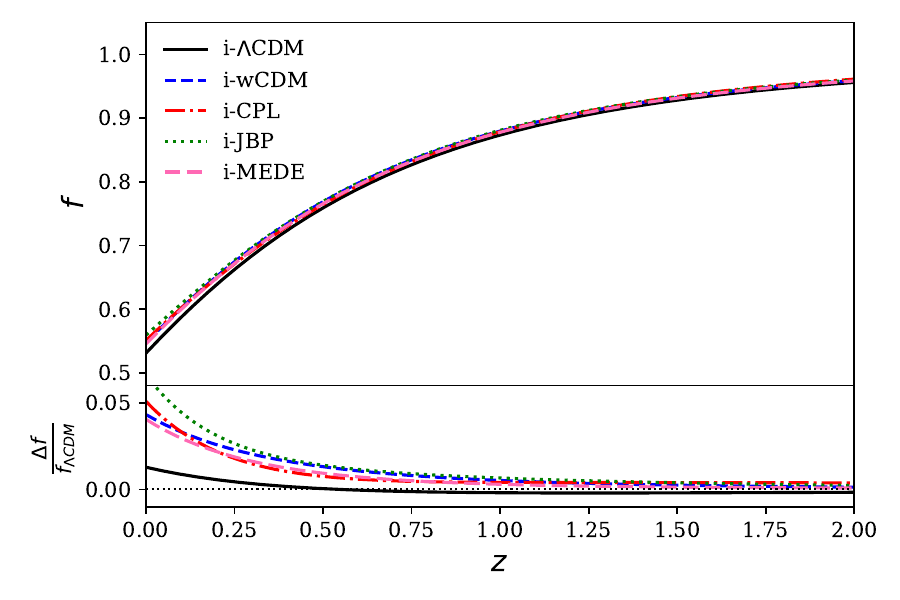}
    \caption{Plot of the redshift evolution of the growth factor for the different iDMDE with i-C/DDE scenarios considered in this work ($c_s^2=1$). Here, $\Delta f = f_{\text{model}} - f_{\Lambda\text{CDM}}$ for the different iDMDE models under consideration.}
    \label{fig:growthfactor}

\end{figure}

\begin{figure*} 
    \centering
    \includegraphics[width=\textwidth]{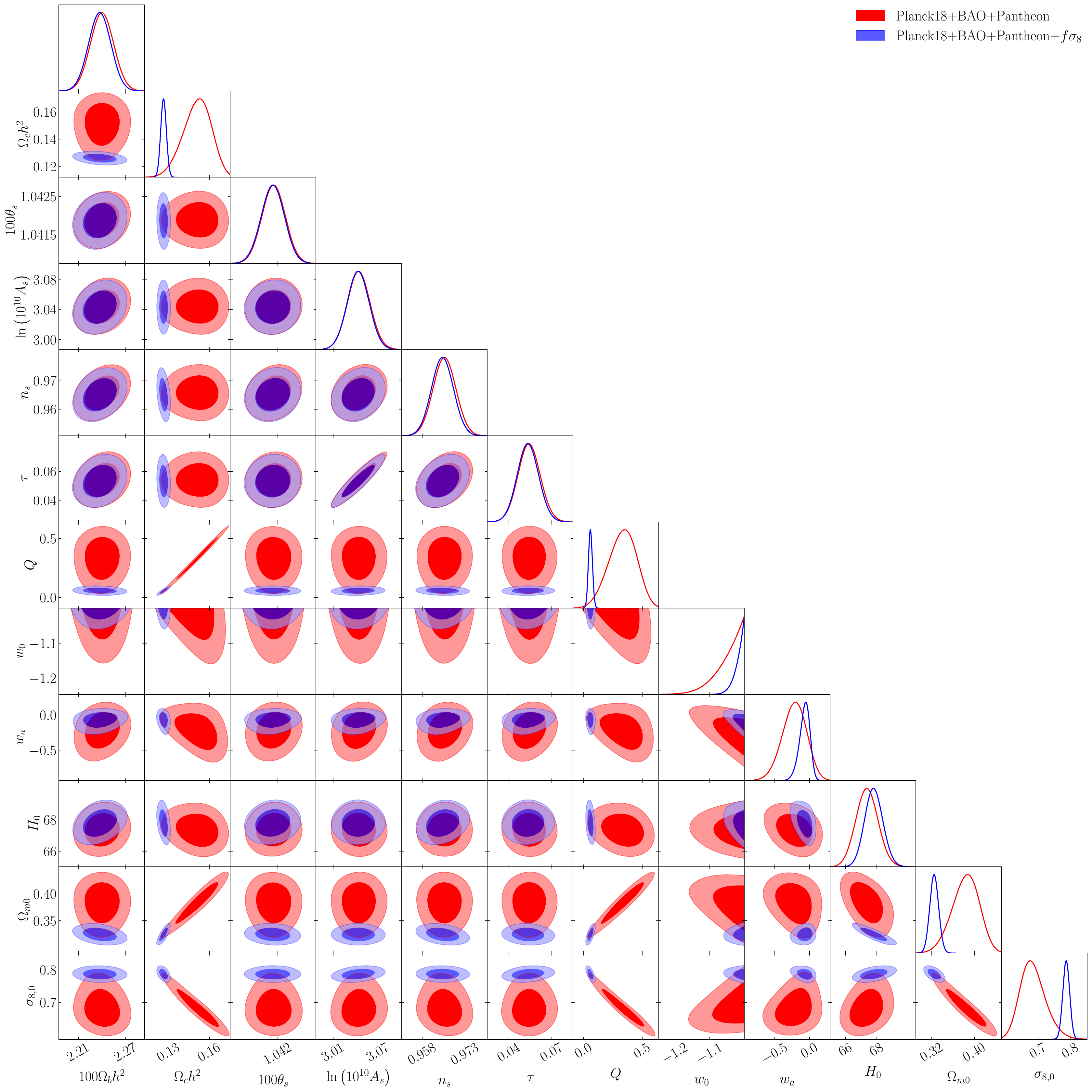}
    \caption{Comparison of constraints obtained for the i-CPL DE model in the phantom regime, for $c_s^2=1$, using combined \textit{Planck} 2018 + BAO + Pantheon+ observational data, versus the addition of $f\sigma_8$ RSD data (compilation taken from \citet{Sinha:2021tnr}).}
    \label{fig:fsigma8compare}
\end{figure*}

The clustering parameter $\sigma_{8,0}$ is related to $f\sigma_{8}$, which is probed by RSD data. Here $f$ is the growth rate of structure formation, which in the interacting dark scenario is given as \citep{Costa:2016tpb}
\begin{equation} \label{eq:f}
    f \equiv \frac{d \ln \delta_m}{d \ln a}=\frac{\mathcal{H}^{-1}}{\delta_m}\left(\frac{\dot{\delta}_c \rho_c+\dot{\delta}_b \rho_b}{\rho_m}+\delta_c \frac{a Q}{\rho_m}-\delta_m \frac{a Q}{\rho_m}\right)\:.
\end{equation}
Here, the evolution of the matter density contrast $\delta_m$, in a linearized approximation, is given by \citep{Dodelson:2003ft}
\begin{equation}\label{eq:growth}
    \delta_m^{\prime \prime}+\left(\frac{3}{a}+\frac{H^{\prime}}{H}\right) \delta_m^{\prime}-\frac{3}{2} \frac{\Omega_{m0}}{a^5 H^2}\delta_m = 0\:\:,
\end{equation}
where $ H = a^{-1} \mathcal{H}$, such that
\begin{equation}
\begin{split}
    H^2 &= H_0^2 \left[\Omega_{m0} \, a^{-3} \, + (1-\Omega_{m0}) \right. \times \\
    &\quad \left. \phantom{(1-\Omega_{m0})} \left( a^{-3} \, Q \int_1^a \frac{a'^2 \, e^{\xi(a')}}{a'^Q} \, \mathrm{d} a' + \frac{e^{\xi(a)}}{a^Q}\right)\right]\:\:,
\end{split}
\end{equation}
and $\xi(a)=\int_{a}^{1} \frac{3[1+w(a')]}{a'}\mathrm{d} a'$. Equation \eqref{eq:f} is a consequence of the non-similar evolution of CDM and baryons in the presence of interactions. The standard scenario is restored for a null interaction term. Observationally, $f$ does not promise to be a very reliable quantity given its dependence on the bias parameter, which relates tracers of matter (such as luminous galaxies) with the underlying total matter distribution ($\delta_g=b\delta_m$). In this regard, a more dependable quantity is the product $f(z)\sigma_{8}(z)$, where $\sigma_8(z)$ is the root mean square of mass fluctuations within a sphere of radius $R=8h^{-1}$ Mpc defined as follows:
\begin{equation}
    \sigma^2(R, z)=\frac{1}{2 \pi^2} \int_0^\infty k^2 P(k, z) W(k R)^2 dk\:,
\end{equation}
where $P(k,z)$ is the power spectrum and $W(kR)$ is a top-hat window function. We proceed to investigate the effects the $f\sigma_8$ data sets from RSD would have on the clustering parameter. We consider the data compiled in \citet{Sinha:2021tnr}. We present the results for the different i-C/DDE cases in Table \ref{tab:fsigma8}. Since we are dealing with C/DDE models, it is imperative to consider the possible non-trivial effects of these models in the growth equation to avoid biases towards $\Lambda$CDM. We carefully address this aspect analytically in our analysis to strive for an unbiased assessment. It should be noted, however, that the RSD data set is extracted by using $\Lambda$CDM as a fiducial model. In this study we focus on examining the role of the existing RSD data for the constraints on the models under consideration here, and we leave a more comprehensive treatment of this possible bias for a future study. In Fig. \ref{fig:growthfactor}, we note that all the interacting C/DDE models prefer a higher value of the growth rate $f$ today, compared to the vanilla $\Lambda$CDM model. This is a consequence of a positive interaction parameter, which indicates a transfer of energy from the DE sector to the DM sector, and hence an increase in clustering in the latter sector.

The comparison between results obtained with and without RSD data for the interacting CPL model is illustrated in Fig. \ref{fig:fsigma8compare}. We see that the addition of the RSD data tends to have improved the constraints on certain parameters, with an order-of-magnitude increase in precision for $\Omega_{m0}$ and $\sigma_{8,0}$.  The interaction parameter $Q$ is also constrained better, with the possibility regarding the presence of a non-zero $Q$ at the 1$\sigma$ confidence level. However, there is no appreciable shift in $H_0$, similar to the previous phantom DE EoS scenarios. The joint constraints with the RSD data prefer higher values of $S_8$, closer to those obtained from KiDS-450+GAMA [$S_8=0.800^{+0.029}_{-0.027}$ \citep{vanUitert:2017ieu}] and HSC SSP [$S_8 = 0.804^{+0.032}_{-0.029}$ \citep{Hamana:2019etx}]. These constraints fall in between the late-time measurements by CFHTLens [$S_8=0.732^{+0.029}_{-0.031}$ \citep{Joudaki:2016mvz}], KiDS-450 [$S_8 = 0.745 \pm 0.039$ \citep{Joudaki:2016kym}] and DES-Y3 [$S_8=0.759^{+0.025}_{-0.023}$ \citep{DES:2021bvc,DES:2021vln}] on one side versus the early Universe CMB \textit{Planck} 2018 constraints [$S_8 = 0.832\pm 0.013$ \citep{Pl18VI}] on the other side, with better precision.

\section{DE interacting with warm DM}\label{sec:wdm}
\begin{table} 
    \resizebox{0.5\textwidth}{!}{\renewcommand{\arraystretch}{1.3} \setlength{\tabcolsep}{18pt} 
    \begin{tabular}{c c c c}
        \hline\hline
        \textbf{Parameters} & \textbf{i-$w$CDM} & \textbf{i-$w$WDM} \\ \hline
        {\boldmath${\Omega_b}{h^2}$} & $0.02243\pm 0.00014        $ & $0.02234\pm 0.00015         $\\

        {\boldmath${\Omega_c}{h^2}$} & $0.143\pm 0.012            $ & $0.1418\pm 0.0095           $\\
        
        {\boldmath$100{\theta_s}$} & $1.04194\pm 0.00029        $ & $1.04191^{+0.00030}_{-0.00027}$\\
        
        {\boldmath${\ln{\left({10^{10}A_s}\right)}}$} & $3.048\pm 0.014 $ & $3.046\pm 0.015       $\\
        
        {\boldmath$n_s$          } & $0.9669\pm 0.0038          $ & $0.9667\pm 0.0038             $\\
        
        {\boldmath${\tau}$       } & $0.0561\pm 0.0074          $ & $0.0547\pm 0.0075             $\\
        
        {\boldmath$Q$            } & $0.25^{+0.13}_{-0.16}      $ & $0.24\pm 0.11                 $\\
        
        {\boldmath$w_0$          } & $-1.066^{+0.061}_{-0.025}  $ & $-1.045^{+0.041}_{-0.017}     $\\
        
        {\boldmath$w_{DM}$       } &              -               & $0.00048^{+0.00017}_{-0.00043}$\\
        
        \hline
        
        {\boldmath$H_0$          } & $67.18\pm 0.67             $ & $67.43\pm 0.64                $\\
        
        {\boldmath$\Omega_{m0}$  } & $0.368\pm 0.030            $ & $0.363^{+0.026}_{-0.022}      $\\
        
        {\boldmath$\sigma_{8,0}$ } & $0.709^{+0.039}_{-0.061}   $ & $0.714^{+0.034}_{-0.048}      $\\
        
        {\boldmath$S_8$ }          & $0.780_{-0.024}^{+0.028}   $ & $0.782_{-0.021}^{+0.023}      $\\
        
        \hline
        \hline
    \end{tabular}
    }
\caption{The mean and 1$\sigma$ constraints obtained for the interacting $w=w_0$ DE model using combined \textit{Planck} 2018 + BAO + Pantheon+ observational data, in the phantom regime, for $c_s^2=1$, drawing a comparison between the i-$w$CDM case and an i-$w$WDM case characterized by an extra EoS parameter $\mathrm{w_{DM}}$.}
\label{tab:wcdmvscdm}
\end{table}

\begin{figure*} 
    \centering
    \includegraphics[width=\textwidth]{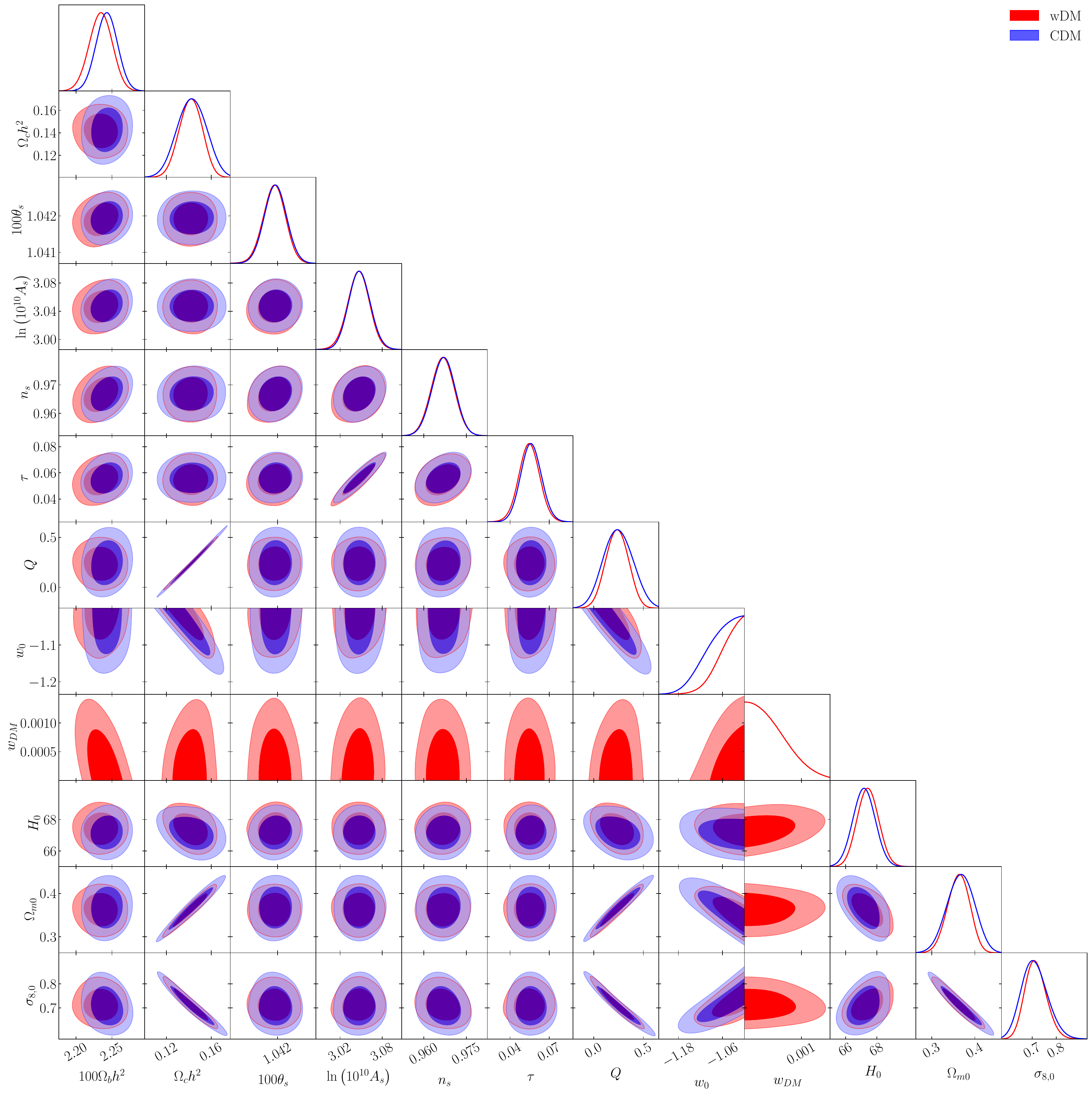}
    \caption{Comparison of constraints obtained for the interacting $w=w_0$ DE model using combined \textit{Planck} 2018 + BAO + Pantheon+ observational data, in the phantom regime, for $c_s^2=1$, drawing a comparison between the i-$w$CDM case and an i-$w$WDM case characterized by an extra EoS parameter $\mathrm{w_{DM}}$.}
    \label{fig:wdmcompare}
\end{figure*}

In all our analyses, we consistently treated DM as completely non-relativistic, characterized by a vanishing EoS. However, given the significance of the $\sigma_{8,0}$ parameter in our investigation, which quantifies perturbations in the matter sector, it is prudent to explore the impact of DM with a non-trivial EoS.  We particularly focus on the WDM models \citep{Hu:2000ke,Wang:2014ina}. These models have emerged as promising avenues for addressing small-scale phenomena associated with the LSS of the Universe, including the core-cusp problem \citep{Oh:2010mc,deBlok:2002vgq}, missing satellites problem \citep{Moore:1999nt,Klypin:1999uc}, and the too-big-to-fail problem \citep{Boylan-Kolchin:2011qkt,Boylan-Kolchin:2011lmk}. WDM models effectively dampen the power spectrum at smaller scales and represent a minimal extension to $\Lambda$CDM, offering a potential solution to these small-scale challenges \citep{Schneider:2013ria,Bohr:2021bdm}. Such an extension involving the cosmological constant in an interacting scenario has been studied in \citet{Pan:2022qrr}. Consequently, we proceed to investigate this set-up within the DE set-up to discern its impact on the $\sigma_{8,0}$ tension. For this, we introduce a free but constant EoS parameter for DM, $\mathrm{w_{DM}}$, together with a $w=w_0$ parametrization for DE (we call this i-$w$WDM, in order to distinguish it from i-$w$CDM). This modifies the continuity equations as \\\\
\begin{equation}
    \dot{\rho}_{\mathrm{dm}} + 3\mathcal{H}(1+\mathrm{w_{DM}})\rho_{\mathrm{dm}} = a^2 \mathcal{Q}^0_{\mathrm{dm}} = a\mathcal{Q}\:,
\end{equation}
\begin{equation}
    \dot{\rho}_{\mathrm{de}} + 3\mathcal{H}(1 + w_0)\rho_{\mathrm{de}} = a^2 \mathcal{Q}^0_{\mathrm{de}} = -a\mathcal{Q}\:,
\end{equation}
the solutions to which are modified as follows:
\begin{equation}
    \rho_{\mathrm{dm}}=\rho_{\mathrm{dm}, 0} \, a^{-3(1+\mathrm{w_{DM}})}+Q \frac{\rho_{\mathrm{de}, 0} \, a^{-3}}{3 w_0 +Q} \left(1-a^{-3w_0 -Q}\right)\:,
\end{equation}
\begin{equation}
    \rho_{\mathrm{de}}=\rho_{\mathrm{de}, 0} \, a^{-3\left(1+w_0+Q/3\right)}\:.
\end{equation}

Correspondingly, the perturbation equations in the synchronous gauge now read
\begin{align}
    &\dot{\delta}_{\mathrm{dm}}=-(1+\mathrm{w_{DM}})\left(\theta_{\mathrm{dm}}+\frac{\dot{h}}{2}\right)+\mathcal{H} Q \frac{\rho_{\mathrm{de}}}{\rho_{\mathrm{dm}}}\left(\delta_{\mathrm{de}}-\delta_{\mathrm{dm}}\right)+Q \frac{\rho_{\mathrm{de}}}{\rho_{\mathrm{dm}}}\times\\ \notag
    &\left(\frac{k v_T}{3}+\frac{\dot{h}}{6}\right)+3\mathcal{H}\mathrm{w_{DM}}\left[\delta_{\mathrm{dm}}+\{3(1+\mathrm{w_{DM}})-Q\frac{\rho_{\mathrm{de}}}{\rho_{\mathrm{dm}}}\}\frac{\mathcal{H}\theta_{\mathrm{dm}}}{k^2}\right]\:,
\end{align}
\begin{equation}
    \dot{\theta}_{\mathrm{dm}}=-\mathcal{H} \theta_{\mathrm{dm}}\left(1+\frac{Q}{1+\mathrm{w_{DM}}} \frac{\rho_{\mathrm{de}}}{\rho_{\mathrm{dm}}}\right)\:,
\end{equation}
\begin{align}
    \dot{\delta}_{\mathrm{de}}=&-(1+w_0)\left(\theta_{\mathrm{de}}+\frac{\dot{h}}{2}\right)- Q\left(\frac{k v_T}{3}+\frac{\dot{h}}{6}\right) \notag\\
    &-3 \mathcal{H}(c_s^2-w_0) \left[\delta_{\mathrm{de}}+\{3(1+w_0)+Q\} \frac{\mathcal{H} \theta_{\mathrm{de}}}{k^2}\right]\:,
\end{align}
\begin{equation}
    \dot{\theta}_{\mathrm{de}}= -\mathcal{H} \left(1-3c_s^2\right) \theta_{\mathrm{de}} + \frac{Q\mathcal{H}\theta_{\mathrm{de}}}{1+w_0}\left(1+c_s^2\right) +\frac{k^2c_s^2}{1+w_0} \delta_{\mathrm{de}}\:,
\end{equation}
with the usual definitions of the terms. We present a comparison between the results of keeping $\mathrm{w_{DM}}$ free versus fixing it to $\mathrm{w_{DM}}=0$ (the non-relativistic CDM scenario), which is analogous to previous cases, in Fig. \ref{fig:wdmcompare} and Table \ref{tab:wcdmvscdm}. The resulting posterior for $\mathrm{w_{DM}}$ shows a peak very close to $\mathrm{w_{DM}}=0$, with a very small upper bound for $\mathrm{w_{DM}}$. This means that the existence of non-cold DM (albeit slight) in the Universe cannot be discarded. However, such a small number has practically no effect on the parameters of interest, and hence a WDM scenario coupled with an interacting DE set-up seems unlikely to be more beneficial than the standard CDM cases when it comes to relaxing cosmological tensions. We expect similar outcomes for the other DE parametrizations considered in this study.

\section{Conclusions}\label{sec:conclusion}
We re-investigate the viability of iDMDE models in addressing the clustering tension employing the latest CMB data from \textit{Planck} 2018, BAO data, and the Pantheon+ SNIa compilation. Adopting an interaction term proportional to the DE density, we examine various parametrizations of the DE EoS for this interacting scenario. We incorporate the full impact of perturbations in both of the dark sectors, assuming that both DM and DE take part in clustering. The analysis has been undertaken for phantom and non-phantom regimes separately due to their distinct theoretical motivations and to avoid phantom-crossing singularities in perturbation equations. We impose no additional constraints on other parameters, facilitating an agnostic data-driven analysis of the phantom and non-phantom sectors. Furthermore, examining both these scenarios independently can help in better understanding of how each different DE model impacts cosmological tensions and observational data, providing a more comprehensive analysis of these iDMDE models. 

Our results indicate an elimination of the correlation between $H_0$ and $\sigma_{8,0}$ for interacting models with constant/dynamic DE EoS. In the phantom scenario, a positive interaction parameter $Q$ (indicating energy flow from DE to DM) is preferred, with $\sigma_{8,0}$ showing a negative correlation with $Q$, thereby helping to reduce $\sigma_{8,0}$. Conversely, in the non-phantom regime, a negative interaction term is preferred. The same correlations observed in the phantom case also exist in the non-phantom scenario, none the less resulting in unrealistically large values of $\sigma_{8,0}$ resulting in an overproduction of structures, that lead to a worsening of the $S_8$ tension compared to i-$\Lambda$CDM, with apparently no shifts in the $H_0$ values. However, the phantom DE models can successfully alleviate the $S_8$ tension without exacerbating the $H_0$ tension. Furthermore, a phantom EoS is favoured as a better fit to the current data sets. These findings point towards the superiority of the phantom models over the non-phantom ones, in the context of interactions in the dark sector and cosmological tensions.

We further investigated the role of the growth of structure ($f\sigma_8$) measurements from RSD data in addition to the aforementioned combination of data sets. We observe tighter constraints on the parameters, especially $\Omega_{m0}$ and $\sigma_{8,0}$, which result in precise measurements of $S_8$. However, the central values show only minor deviations from the best-fitting values of the concordance $\Lambda$CDM model. Consequently, there is a relatively less pronounced relaxation of the $S_8$ tension. This may suggest an intrinsic bias towards $\Lambda$CDM within this data set, where the data extraction involves fixing the shape of the linear matter power spectrum under a fiducial cosmology. This highlights the need for further investigation into potential underlying assumptions.

To examine any effect of the sound speed of DE ($c_s^2$) in this analysis, we reanalyse the iDMDE models keeping $c_s^2$ as a free parameter to be directly constrained from data. However, our findings indicate that the inclusion of $c_s^2$ as a free parameter has a negligible effect on overall constraints, suggesting that the current data sets are not sensitive to this parameter. This implies the robustness of our results subject to a wide range of possible $c_s^2$ values. Additionally, we also investigate a scenario involving WDM with a non-trivial DM EoS as an extension to the interacting CDM-DE set-up. Our results strongly favour the presence of CDM viz. $\mathrm{w_{DM}}=0$, with an upper bound obtained on $\mathrm{w_{DM}}$ in the interacting set-up with a constant DE EoS. There are minimal shifts observed in constraints on the other parameters, including $H_0$ and $S_8$. So, this interacting WDM-DE set-up has practically no effect on the parameters of interest. Thus, a WDM scenario coupled with an interacting DE set-up seems unlikely to be more beneficial than the standard CDM cases when it comes to relaxing cosmological tensions.

It deserves mention that this whole exercise has been undertaken for the specific parametric form of the interaction term $\mathcal{Q} = H Q \rho_{\mathrm{de}}\:$, where $Q$ represents the coupling or strength of the interaction, as this particular form has previously shown promise in addressing the $H_0$ tension to some extent. However, open questions remain regarding the nature of the interaction term. The possibility of this $Q$ evolving with redshift has been explored in \citet{Sinha:2021tnr}, which calls for further investigation. As an alternative approach, one can attempt a non-parametric reconstruction of $\mathcal{Q}$, where no prior dependence on the functional form exists. Some preliminary investigations have already begun in this direction \citep{Mukherjee:2021ggf, Escamilla:2023shf, Bonilla:2021dql}. However, a comprehensive analysis incorporating different combinations of data sets, and their impact on the cosmological tensions, is yet lacking, which we intend to investigate in future works.

Additionally, there are concerns regarding BAO data, which rely on assuming a fiducial cosmology (typically the $\Lambda$CDM model) when extracting the BAO signal. This reliance on the fiducial template could introduce bias in the resulting parameter constraints for interacting models that significantly deviate from $\Lambda$CDM. Similarly, both RSD data, and the inference of $S_8/\sigma_{8,0}$ from LSS observations, entail some fiducial model dependence. Despite no specific model assumption with regard to the SNIa data, our analysis does assume that the intrinsic luminosities of SNIa do not evolve with redshift. This assumption of the constancy of the SNIa absolute magnitude $M_B$, a topic that has been widely debated \citep{Kang:2019azh}, can influence the constraints on other parameters during likelihood computation, potentially leading to significant differences in results. Investigating the applicability of these data sets to iDMDE models is crucial \citep{DiValentino:2019ffd}, especially given the uncertainty around the physical nature of the interaction parameter. These issues call for further exploration in future. 

In conclusion, the introduction of interaction within the cosmic dark sector effectively eliminates the strong positive correlation existing between $H_0$ and $\sigma_{8,0}$ for non-interacting DM-DE models. A phantom EoS contributes to a reduction in the value of $\sigma_{8,0}$, thereby easing the clustering tension. Notably, the interacting CPL and JBP models, incorporating a phantom DE component, demonstrate the potential to alleviate the clustering tension to $<1\sigma$ without exacerbating the Hubble tension. These results are intriguing as they suggest the potential to address the $S_8$ tension without worsening the $H_0$ tension within the same theoretical framework. We hope to explore further in this direction in the future, considering more DE parametrizations with different combinations of data sets, and investigating the role that DE perturbations could play, in light of the recent Dark Energy Spectroscopic Instrument (DESI) \citep{DESI:2024mwx} data release, as well as upcoming future missions.

\section*{Acknowledgements}
We gratefully acknowledge the use of the publicly available codes \href{https://github.com/lesgourg/class_public}{\textit{CLASS}}, \href{https://github.com/brinckmann/montepython_public}{\textit{MontePython}} and \href{https://github.com/cmbant/getdist}{\textit{GetDist}}. We thank Supriya Pan, Debarun Paul, Antara Dey, Raj Kumar Das, and Swati Gavas for insightful discussions, and Arko Bhaumik and Sourav Pal for invaluable support throughout the duration of this work. We also thank the anonymous reviewer for their valuable suggestions towards the improvement of the manuscript. RS thanks the Indian Statistical Institute (ISI) Kolkata for financial support through Senior Research Fellowship. PM acknowledges the Anusandhan National Research Foundation (ANRF),  Govt. of India for financial support under the National Post-Doctoral Fellowship (N-PDF File no. PDF/2023/001986). SP thanks the Department of Science and Technology, Govt. of India for partial support through grant no. NMICPS/006/MD/2020-21 and also the Anusandhan National Research Foundation (ANRF),  Govt. of India for partial support through project no. CRG/2023/003984. We acknowledge the computational facilities of ISI Kolkata, those made available by the Technology Innovation Hub, ISI Kolkata, and the use of the Pegasus cluster of the high performance computing (HPC) facility at the Inter-University Centre for Astronomy and Astrophysics (IUCAA), Pune, India.

\vspace{-14pt}

\section*{Data Availability}
The data sets used in this work are all publicly available. The modified codes used for this study may be made available upon reasonable request.

\vspace{-14pt}


\bibliographystyle{mnras}
\bibliography{mnras}

\bsp
\label{lastpage}

\end{document}